\documentclass[aps,prl,twocolumn,superscriptaddress]{revtex4-1}
\usepackage{graphicx}
\usepackage{subfigure}
\usepackage{epstopdf}
\usepackage{amsmath}
\usepackage{amssymb}
\usepackage{amsfonts}
\usepackage{mathrsfs}
\usepackage{theorem}
\usepackage{bm}
\usepackage{url}
\usepackage[T1]{fontenc}
\usepackage{csquotes}
\MakeOuterQuote{"}

\usepackage{algorithm}
\usepackage{algorithmicx}
\usepackage{algpseudocode}

\usepackage{dcolumn}
\usepackage{color}

\def\vec#1{\bm{#1}} 

\newcommand{\rmi}{\mathrm{i}}

\newcommand{\ketbra}[2]{\ensuremath{\left| #1 \rangle \langle #2\right|}}

\newcommand{\be}{\begin{equation}}
\newcommand{\ee}{\end{equation}}
\newcommand{\ba}{\begin{align}}
\newcommand{\ea}{\end{align}}

\def\<{\langle}  
\def\>{\rangle}  
\newcommand{\ket}[1]{| #1\>}
\newcommand{\bra}[1]{\< #1|}

\def\eqref#1{\textup{(\ref{#1})}}  
\newcommand{\eref}[1]{Eq.~\textup{(\ref{#1})}}

\newcommand{\fref}[1]{Fig.~\ref{#1}}

\newcommand{\cref}[1]{Conjecture~\ref{#1}}
\newcommand{\Cref}[1]{Conjecture~\ref{#1}}

\begin{document}

\widetext


\title{Control-enhanced sequential scheme for general quantum parameter estimation at the Heisenberg limit}
\author{Zhibo Hou}
\affiliation{Key Laboratory of Quantum Information,University of Science and Technology of China, CAS, Hefei 230026, P. R. China}
\affiliation{CAS Center For Excellence in Quantum Information and Quantum Physics}
\author{Rui-Jia Wang}
\affiliation{Key Laboratory of Quantum Information,University of Science and Technology of China, CAS, Hefei 230026, P. R. China}
\affiliation{CAS Center For Excellence in Quantum Information and Quantum Physics}
\author{Jun-Feng Tang}
\affiliation{Key Laboratory of Quantum Information,University of Science and Technology of China, CAS, Hefei 230026, P. R. China}
\affiliation{CAS Center For Excellence in Quantum Information and Quantum Physics}

\author{Haidong Yuan}
\email{hdyuan@mae.cuhk.edu.hk}
\affiliation{Department of Mechanical and Automation Engineering, The Chinese University of Hong Kong, Shatin, Hong Kong}
\author{Guo-Yong Xiang}
\email{gyxiang@ustc.edu.cn}
\affiliation{Key Laboratory of Quantum Information,University of Science and Technology of China, CAS, Hefei 230026, P. R. China}
\affiliation{CAS Center For Excellence in Quantum Information and Quantum Physics}
\author{Chuan-Feng Li}
\affiliation{Key Laboratory of Quantum Information,University of Science and Technology of China, CAS, Hefei 230026, P. R. China}
\affiliation{CAS Center For Excellence in Quantum Information and Quantum Physics}
\author{Guang-Can Guo}
\affiliation{Key Laboratory of Quantum Information,University of Science and Technology of China, CAS, Hefei 230026, P. R. China}
\affiliation{CAS Center For Excellence in Quantum Information and Quantum Physics}


\date{\today}






\begin{abstract}
The advantage of quantum metrology has been experimentally demonstrated for phase estimations where the dynamics are commuting. General noncommuting dynamics, however, can have distinct features. For example, the direct sequential scheme, which can achieve the Heisenberg scaling for the phase estimation under commuting dynamics, can have even worse performances than the classical scheme under noncommuting dynamics. Here we realize a scalable optimally controlled sequential scheme, which can achieve the Heisenberg precision under general noncommuting dynamics. We also present an intuitive geometrical framework for the controlled scheme and identify sweet spots in time at which the optimal controls used in the scheme can be pre-fixed without adaptation, which simplifies the experimental protocols significantly. We successfully implement the scheme up to eight controls in an optical platform, demonstrate a precision near the Heisenberg limit. Our work opens the avenue for harvesting the power of quantum control in quantum metrology, and provides a control-enhanced recipe to achieve the Heisenberg precision under general noncommuting dynamics.

\end{abstract}

\maketitle


\emph{Introduction.}--Improving the measurement precision \cite{GiovLM04,Naga07beating,HiggBBW07,GiovLM11,XianHBW11,Slus17unconditional} is one of the major driving forces for technology and science. The precision of a measurement scheme is ultimately bounded by the available resources \cite{caves1981quantum,yurke19862}, which are typically quantified by the number of uses of a discrete-time dynamics, $N$, or by the evolution time of of a continuous-time dynamics, $T$. The best precision of a classical scheme, known as the quantum Shot-Noise-Limit (SNL), scales as $1/\sqrt{N}$ or $1/\sqrt{T}$ for discrete and continuous dynamics respectively. The SNL is already constraining the performance of current state-of-art precision measurements, such as LIGO interferometer \cite{caves1981quantum,schnabel2010quantum,abadie2011gravitational,aasi2013enhanced}. By exploring quantum effects, quantum metrology can surpass the SNL \cite{Mitchell04super,walther2004broglie,GiovLM04,GiovLM06,GiovLM11}. For example, by preparing the probe state as the NOON state in the entangled parallel scheme \cite{Boll96optimal,Lee02a}, it can achieve the Heisenberg precision which scales as $1/{N}$\cite{GiovLM04,Naga07beating,Okam08beating,GiovLM11,XianHBW11}.
In practise, however, preparing large entangled states are extremely challenging. To date, the largest NOON state prepared deterministically in optical experiment is $N=5$ \cite{Afek10high}, while the largest NOON state that has been implemented for quantum metrology is only $N=4$(2) with(without) the postselection \cite{Naga07beating,Okam08beating,XianHBW11}. 

\begin{figure}[t]
	\center{\includegraphics[width=0.46\textwidth]{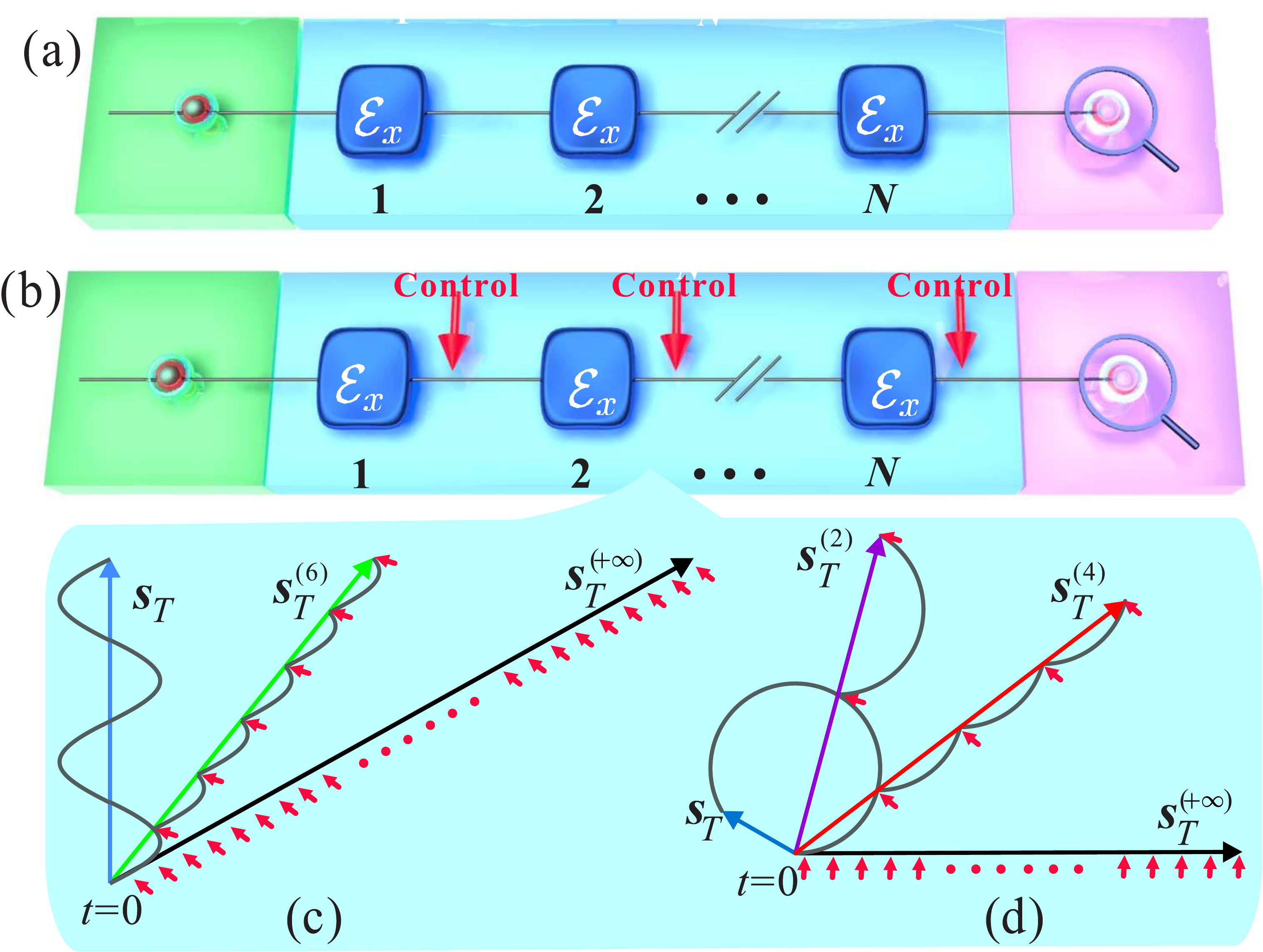}}
	\caption{\label{fig:scheme} {Generators in the direct and control-enhanced sequential schemes.} (a)  Direct sequential scheme; (b) Control-enhanced sequential scheme with $N$ controls. Both schemes consist of three steps: state preparation(green module), evolution(blue) and measurement(purple). (c) The evolution of the Bloch vectors $\vec{s}_T^{(N)}$ for the generators ${S}_T^{(N)}$, here $N$ denotes the number of controls. Without the controls the path of $\vec{s}_T$ is a helical line. The controls change the velocity and increase the length of $\vec{s}_T^{(N)}$.  (d) The evolution of the generators at the plane orthogonal to $\vec{n}_h$. Without the controls $\vec{s}_T$ has a uniform circular motion at the plane orthogonal to $\vec{n}_h$. The controls change the motion and increase the length, as shown by the paths of $\vec{s}_T^{(N)}$.}
\end{figure}

The direct sequential scheme, in which the probe state evolves under the same dynamic multiple times as shown in \fref{fig:scheme}(a), can achieve the same (Heisenberg) precision if the dynamics commute with each other at different values of the parameter. For example, for the usual phase estimation, the dynamic is given by $U=e^{-\rmi\phi H}$, which commute with each other at different values of $\phi$, the direct sequential scheme can then achieve the Heisenberg precision for the usual phase estimation\cite{GiovLM06,GiovLM11,Berr09how}. The sequential scheme is practically more scalable as entanglement is not necessarily required. Higgins \textit{et al.} \cite{HiggBBW07} have experimentally implemented the direct sequential scheme for the estimation of an optical phase, demonstrate a Heisenberg-limited precision. 

The direct sequential scheme, however, can not achieve the Heisenberg precision under general dynamics that do not commute at different values of the parameter (which we will refer as noncommuting dynamics). It can have even worse performances than the shot-noise limit\cite{pang2014}. \citet{Haid15optimal} showed the Heisenberg precision can be restored under general noncommuting dynamics by adding additional quantum controls as in \fref{fig:scheme}(b). Here we first investigated this control-enhanced sequential scheme in a more intuitive and geometrical way to show how the generator of the parameter is coherently accumulated and how quantum control can increase the norm of the generator. We then identify some sweet spots in time at which the optimal controls can be pre-fixed without the need of adaptation. This simplifies the practical implementations significantly. We then experimentally implement the control-enhanced scheme and achieve a precision near the Heisenberg limit for the estimation of the orientation of an optical plate whose dynamics is non-commuting. 

\emph{Control-enhanced sequential scheme for general dynamics.}--The precision limit in quantum metrology can be calibrated by the quantum Cramer-Rao bound(QCRB)\cite{GiovLM11,Hole82book,Hels76book} as  $\delta\hat{\phi}\geq \frac{1}{\sqrt{nJ}}$, where $\delta\hat{\phi}=\sqrt{E[(\hat{\phi}-\phi)^2]}$ is the standard deviation of an unbiased estimator, $n$ is the number of times the measurement is repeated, $J$ is the Quantum Fisher Information(QFI) which bounds the precision limit\cite{Hels76book}. 

For a general unitary dynamics, $U_t(\phi)=e^{-iH(\phi)t}$,  QFI is determined by the variance of its generator as
\begin{equation}\label{key}
J=4\<\Delta S_T^2\>
\end{equation}
where the generator $S_T$ is defined as $S_T=\rmi[\partial_\phi U_T(\phi)]U^\dagger_T(\phi)$ and $\<\Delta S_T^2\>=\bra{\psi} S_T^2\ket{\psi}-\bra{\psi} S_T\ket{\psi}^2$\cite{pang2014}. For general time-independent Hamiltonian we have \cite{pang2017optimal}
\begin{equation}\label{eq:generator}
S_T=\int_{0}^{T}V_{t}d{t},
\end{equation}
where $V_{t}=U^\dagger_{t}(\phi) V_0 U_{t}(\phi)$ with $V_0=\partial_\phi H(\phi)$. $S_T$ can be regarded as an overall signal strength which is coherently accumulated from the instantaneous signal $V_{t}$ over a period of time. For commuting dynamics, where $[U_t(\phi),U_t(\phi+d\phi)]=0$, we have $[H(\phi),\partial_\phi H(\phi)]=0$, $V_{t}=V_0$. In this case $S_T=V_0T$, $\<\Delta S_T^2\>=T^2\<\Delta V_0^2\>$, which scales as $T^2$ and leads to the Heisenberg limit.

For noncommuting dynamics, however, things are more different. We consider a general Hamiltonian $H$ on a qubit, which can be written as $H=\vec{h}\cdot\vec{\sigma}$, $\vec{h}=(h_1,h_2,h_3)$ and $\vec{\sigma}=(\sigma_1,\sigma_2,\sigma_3),$ where
\[\sigma_1=\left(
\begin{array}{cc}
0 & 1 \\
1 & 0 \\
\end{array}
\right),\quad
\sigma_2=\left(
\begin{array}{cc}
0 & -\rmi \\
\rmi & 0 \\
\end{array}
\right),\quad
\sigma_3=\left(
\begin{array}{cc}
1 & 0 \\
0 & -1 \\
\end{array}
\right)\] are Pauli matrices. Similarly we can write $S_T=\vec{s}_T \cdot\vec{\sigma}$ and $V_t=\vec{v}_t \cdot\vec{\sigma}$. The Bloch vectors $\vec{s}_T$ and $\vec{v}_t$ can be regarded as the displacement and the velocity respectively. As shown in \fref{fig:scheme}(c)(see Supplemental Material for derivation),  the trajectory of $\vec{s}_T$ is a helical line, its parallel component along $\vec{h}$ is a uniform rectilinear motion with the speed $\vec{v}_0\cdot\vec{n}_h$, $\vec{n}_h=\vec{h}/|\vec{h}|$, its perpendicular component has a circular motion with the radius $\frac{\sqrt{\left|\vec{v}_0\right|^2-(\vec{v}_0\cdot\vec{n}_{h})^2}}{2|\vec{h}|}$ and the angular speed $2|\vec{h}|$. The variance of $S_T$, $\<\Delta S_T^2\>=|\vec{s}_T|^2-\bra{\psi}S_T\ket{\psi}^2$, is upper bounded by (see Supplemental Material)
\begin{equation}
|\vec{s}_T|^2=\left(\vec{v}_0\cdot\vec{n}_h \right)^2T^2+\frac{\left|\vec{v}_0\right|^2-(\vec{v}_0\cdot\vec{n}_{h})^2}{|\vec{h}|^2}\sin^2 |\vec{h}T|,
\end{equation}
 which is smaller than $\left|\vec{v}_0\right|^2T^2$.

However, if additional controls are available, we can use the controls to change the velocity and increase the length of the generator as shown in \fref{fig:scheme}(c,d). Under such control-enhanced sequential scheme the total dynamics is given by $U_{T}^{(N)}(\phi)=U_{ct}^N$ with $U_{ct}=U_cU_{t}(\phi)$, here $t=T/N$ and $U_c$ is the added control after each evolution of time $t$. The generator for this controlled dynamics at time $T$ is
\[S_{T}^{(N)}=\rmi \left[U_{ct}^N\right]^\dagger\partial_x U_{ct}^N=\sum\limits_{k=0}^{N-1} \left[U_{ct}^{k}\right]^\dagger S_{t}U_{ct}^{k},\]
where we use $S_{T}^{(N)}$ to denote the generator after adding $N$ controls and $S_T$ as the generator of the free evolution. When $N=1$, i.e., no controls added during the evolution(only one control at the end of the evolution which does not change the QFI), $t=T/N=T$, $S_{T}^{(1)}=S_T$ which leads to the result in \eref{eq:generator}. With general $N$ controls, to maximize the variance of $S_{T}^{(N)}$, we can choose $U_c=e^{-\rmi\alpha\vec{s}_{t}}U_{t}^\dagger(\phi)$, where $\alpha$ can be chosen arbitrarily and is typically set as $0$. In this case $[U_{ct},S_{t}]=0$ and $S_{T}^{(N)}=NS_{t}=NS_{T/N}$. The variance of the generator is then
\begin{eqnarray}
\aligned
&\<\Delta (S_T^{(N)})^2\>=N^2\<\Delta S_t^2\>\\
=&N^2\left[\left(\vec{v}_0\cdot\vec{n}_h\right)^2\frac{T^2}{N^2}+\frac{\left|\vec{v}_0\right|^2-(\vec{v}_0\cdot\vec{n}_{h})^2}{|\vec{h}|^2}\sin^2 |\vec{h}\frac{T}{N}|\right].
\endaligned
\end{eqnarray}
When $N\to\infty$, this goes to $\left|\vec{v}_0\right|^2T^2$ which restores the Heisenberg limit.

\begin{figure*}[t]
	\center{\includegraphics[width=1\textwidth]{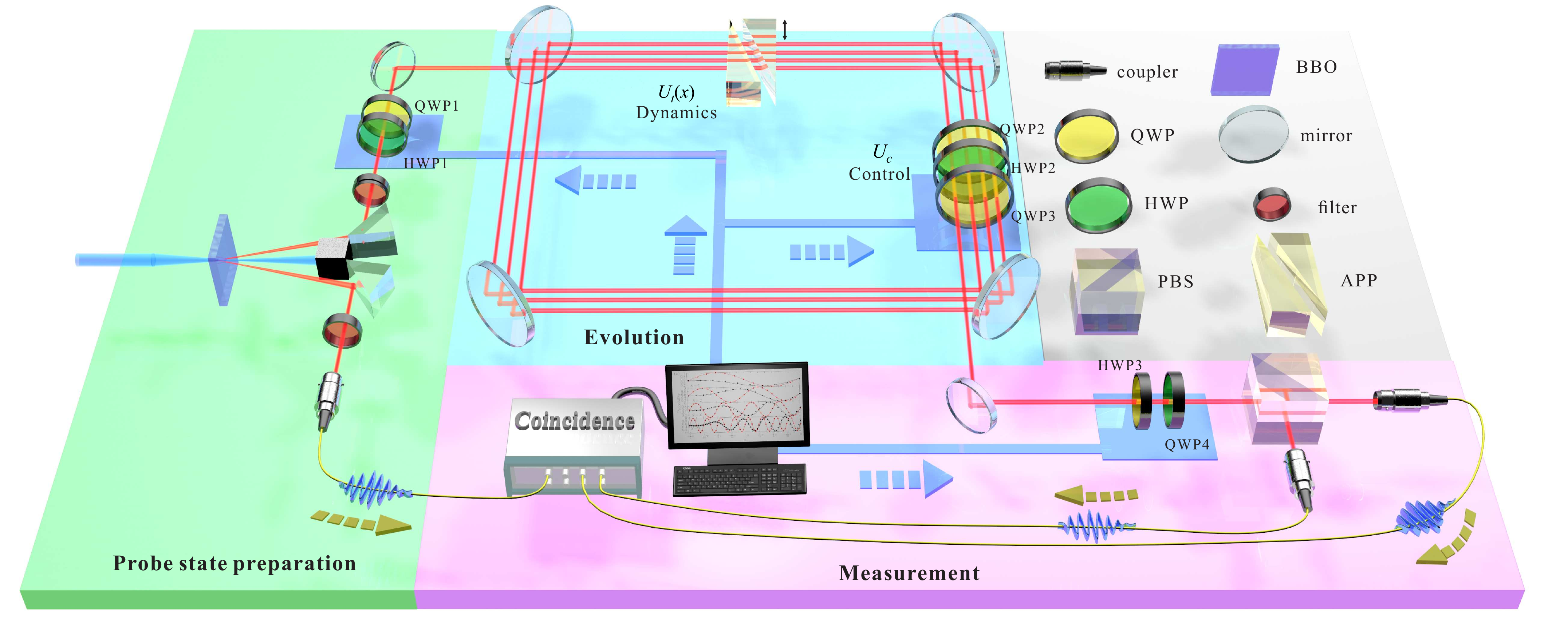}}
	\caption{\label{fig:setup}{Experimental setup.} The module of preparation prepares the probe state using the polarization degree of a heralded single photon from SPDC process. The probe state then undergoes the evolution and the control in the module of evolution. The state is then measured in the module of measurement. Key devices in the setup: BBO--$\beta$-barium-borate crystal, QWP--quarter-wave plate,  HWP--half-wave plate, PBS--polarizing beam splitter, APP--adjustable phase plate.}
\end{figure*}

\emph{Experimental setup.}--Our control-enhanced experiment has three modules: preparation, evolution and measurement, as shown in \fref{fig:setup}. In the preparation module, a 1-mm-long $\beta$-barium-borate(BBO) crystal crystal, cut for type-\uppercase\expandafter{\romannumeral1} phase-matched spontaneous parametric down-conversion (SPDC) process, is pumped by a 40-mW horizontally polarized beam at 404~nm to generate heralded single photons at the rate of 3500 Hz \cite{Kwia99ultrabright}. We then use a combination of a half-wave plate(HWP) and a quarter-wave plate(QWP) to prepare the photon in any desired polarization, which is used as the probe state.  In the evolution module, we use an adjustable phase plate(APP), which is realized with a Soleil-Babinet Compensator, to generate the noncommuting dynamics on the polarization of the photon. When the optic axis of APP is deviated from the horizontal direction by an angle, $x$, the two polarization states, in the basis of the horizontal and vertical polarization with $\ket{0}=\ket{H}$ and $\ket{1}=\ket{V}$, can be written as $\ket{o}=\cos x\ket{0}+\sin x\ket{1}$ and $\ket{e}=-\sin x\ket{0}+\cos x\ket{1}.$
When a photon passes through the phase plate with a $2t$-phase shift, it undergoes a unitary evolution
$U_t(x)=\ketbra{o}{o}+e^{\rmi2t}\ketbra{e}{e},$
which can be rewritten as
$U_t(x)=e^{-\rmi(\sin2x\sigma_1+\cos2x\sigma_3)t}$ in the basis of the horizontal and vertical polarization.
By controlling the phase $t$, this is equivalent to a time evolution governed by the Hamiltonian $H=\sin2x\sigma_1+\cos2x\sigma_3$, here the parameter $x$ represents the angle between the optical axis of the phase plate and the horizontal direction. The estimation of $x$ thus corresponds to the estimation of the orientation of the phase plate. The control is realized by a combination of two QWPs and a HWP, which is capable of generating arbitrary unitary operation on the polarization. Multiple passes of the qubit are realized by a cavity loop made of four mirrors. The number of controls is deterministically controlled by moving the translation stage of one mirror, which can be realized without affecting the coupling efficiency in the measurement module (see Supplemental material). The module of measurement consists of the HWP, QWP, PBS and two single-photon detectors which can perform the projective measurements along any desired direction.

\emph{Pre-fixed control at the sweet spots in time.}--The optimal control in general depends on the parameter and can only be realized adaptively, but there are some cases the adaptation is not required. In our experiment, the noncommuting dynamics is governed by the Hamiltonian $H(x)=\sin2x\sigma_1+\cos2x\sigma_3$. It is easy to obtain $V_0=\partial_xH(x)=2(\cos2x,0,-\sin2x)\cdot \vec{\sigma}$. The vector, $\vec{v}_0=2(\cos2x,0,-\sin2x)$, is orthogonal to the Hamiltonian vector $\vec{n}_h=(\sin2x,0,\cos2x)$, where $H=\vec{n}_h\cdot\vec{\sigma}$. Thus, without controls the generator only has a perpendicular component in a circular motion as shown in \fref{fig:scheme}(d). The largest variance of $S_T$ is $4\sin^2T$ at time $T$, which is much lower than the Heisenberg limit $4T^2$ \cite{Haid15optimal}.

Under the control-enhanced sequential scheme with $N$ passes through the dynamics and control, the QFI can reach $16N^2\sin^2t$. In real experiments, the number of controls are always limited. The maximal QFI that can be achieved with $N$ controls is $16N^2$, where the minimal $t$ attaining this maximal value is $t=\frac{\pi}{2}$. Under $N$ controls, $T=Nt=\frac{\pi}{2}N$ is the smallest total time to achieve the maximal value. In addition, at these time points the optimal control can all be taken as $U_c=\rmi \sigma_3$, which is independent of $x$ and can be prefixed without adaptation. This control works for all $x$ at $t=\frac{\pi}{2}$, as $U_{ct}=\rmi \sigma_3e^{-i\frac{\pi}{2}H(x)}=e^{\rmi  2x\sigma_2}$ commute with $S_t=2\sigma_2$ for all $x$. Thus when $N$ controls are used, at $T=\frac{\pi}{2}N$, the QFI can achieve the maximal value $J_T^{(N)}=16N^2$ with the pre-fixed control $U_c=\rmi \sigma_3$.

\begin{figure}[t]
	\center{\includegraphics[scale=0.65]{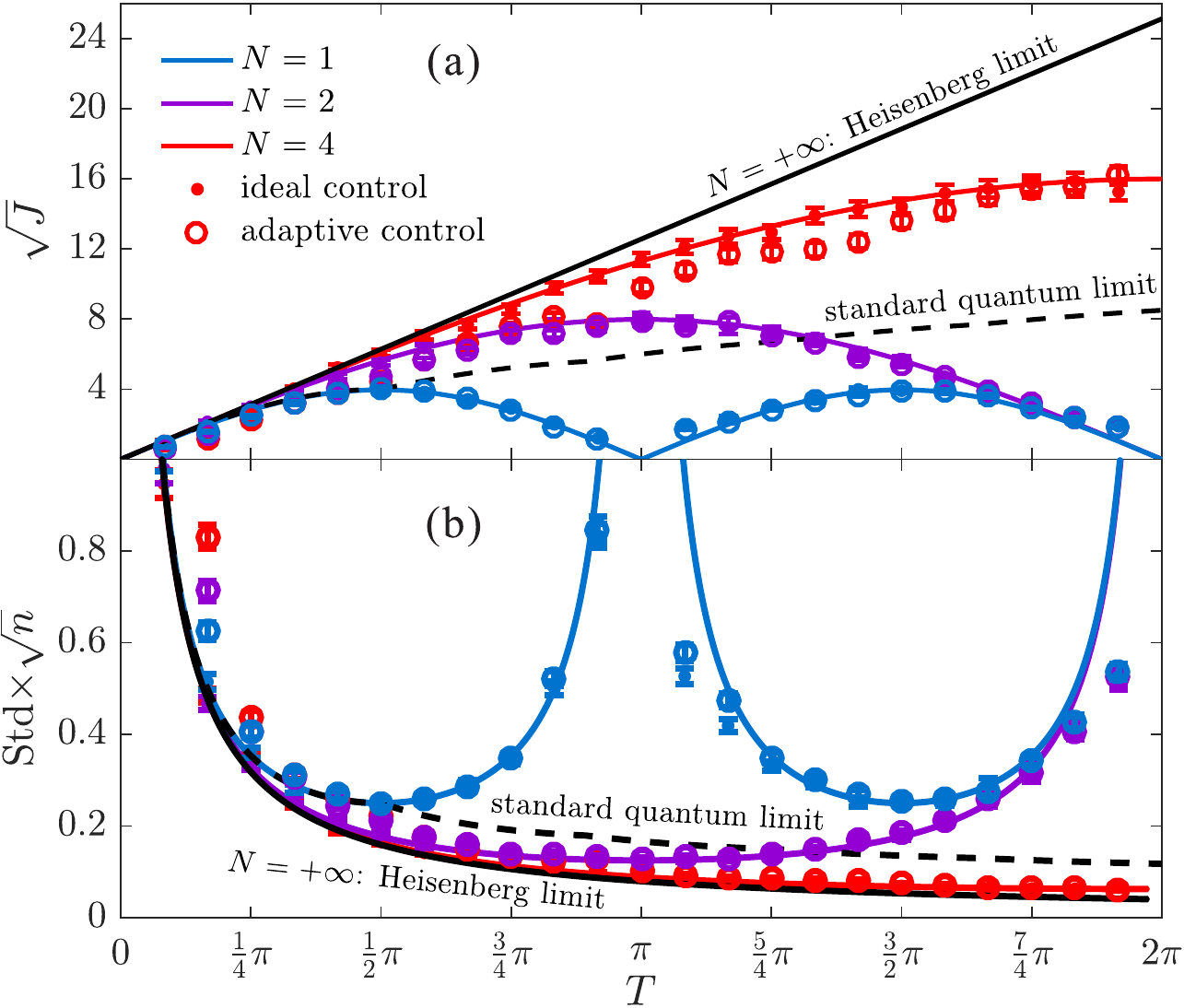}}
	\caption{\label{fig:adaptive}{Precision with the optimal and adaptive controls.} (a) QFI; (b) The standard deviation; The performances with $N=1, 2$ and $4$ controls are demonstrated, which are denoted by blue, purple and red colors, respectively.  Experimental results for ideal controls (dots) and adaptive controls (circles)  are close to optimal theoretical values (solid lines). The error bars are discussed in Supplementary material.}
\end{figure}

\begin{figure}[t]
	\center{\includegraphics[scale=0.5]{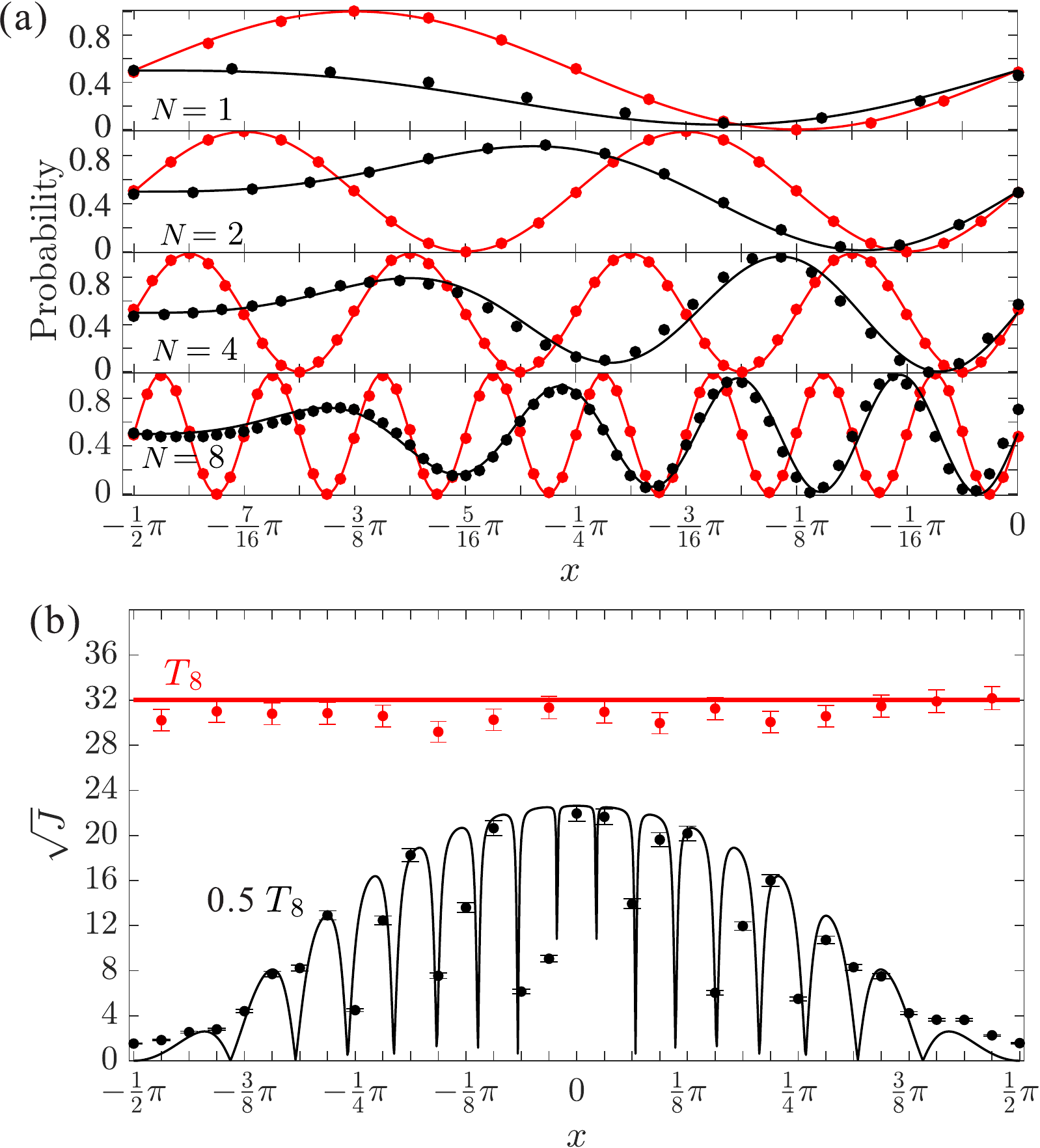}}
	\caption{\label{fig:probability exp}{Experimental results at the sweet spots in time}. (a)Probability distribution with respect to $x$. Red and black dots show frequencies measured experimentally with 50000 measurements at sweet time $t$ and non-sweet time $0.5t$, respectively. The solid lines show the theoretical probability distribution. From upper to bottom, the four subplots correspond to control number $N=1,2,4$ and 8. Error bars are calculated from measurement statistics and too small to be visible. (b)  The QFI for the case of $N=8$ is plotted, at both the sweet spot in time $T_8=4\pi$ and the non-sweet spot in time $0.5T_8=2\pi$. The solid lines are theoretical value and dots are experiment results.}
\end{figure}

\emph{Experimental results at any given time with different number of controls.}--In the first set of experiments, we demonstrate the precision scaling with respect to the evolution time $T$ when different number of controls are used. For any given $T$, if $N$ controls are used ($t=\frac{T}{N}$), the QFI under the control-enhanced scheme can reach $J_{T}^{(N)}=N^2J_{t}=16N^2\sin^2\frac{T}{N}$, which increases with $N$.

We first consider the scenario when $x$ is known to be within a very small neighborhood so that we can choose ideal controls. In the experiment $x$ is close to $0$ and the optimal probe states, controls and measurements are prepared according to $x=0$ and $t=T/N$ (see detailed experimental implementation in Supplementary material). 
We make $n(=50)$ measurements to get the probabilities of the two outcomes. To get the statistics of $\hat{x}$ experimentally, we repeat the process 1000 times to get the distribution of $\hat{x}$, from which the standard deviation of the estimator, $\delta\hat{x}$, is obtained. As shown in \fref{fig:adaptive}(a), the experimental precision (dots) saturates the theoretical optimal value. It can also be seen that when the number of controls increases from 1 to 4, the precision beats the shot noise limit (see Supplemental material) and gets closer to the Heisenberg limit.

In the second scenario, $x$ can be any value within an interval, where the size of the interval is only restricted by phase ambiguities\cite{HiggBBW07,Berr09how} (see Supplemental material). In this case we used adaptive controls.  For each round we make $5$ iterations of the adaptation. Specifically, we make new estimations of the parameter after each $10$ measurements.  
The controls in the first $10$ measurements are designed according to $\hat{x}=\pi/4$, the middle point of $[0,\frac{\pi}{2}]$, as $U_c=U_t^\dagger(\frac{\pi}{4})$, then are adaptively updated (see Supplementary material for experimental implementation) based on the new estimated value $\hat{x}$ obtained with the maximum likelihood estimation, which maximizes the posterior probability based on the obtained data. 
In \fref{fig:adaptive}, we plotted the precisions (circles) achieved by the adaptively controlled scheme. It can be seen that for $N=1$ and $2$, the obtained precision is almost the same as the theoretical optimal value; for $N=4$, the precision is slightly smaller, but already quite close to the the theoretical optimal value, i.e., the adaptive controls are already close to be optimal after five iterations. The results also clearly show that the precisions beat the shot noise limit and get closer to the Heisenberg limit.

\emph{Experimental results with a given number of controls at the sweet spots in time.}--In the second set of experiments, we carry out the experiments under any given number of controls and show the advantages at the sweet spots in time. At general time points, the  controls typically depend on the actual value of the parameter, thus need to be updated adaptively. With a given number of controls, at the sweet spots in time they can all be pre-fixed. Specifically, the optimal probe state at the sweet spot in time is $\ket{\psi}=\ket{H}$, the optimal control is $U_c=\rmi\sigma_3$ and the optimal measurement is the projective measurements on the eigenvectors of $\sigma_1$. They are all independent of the actual value of the parameter.
If $N$ controls are used, then the probabilities of the two measurement outcomes are $\frac{1\pm\cos4Nx}{2}$ (see Supplementart material). We plot the probability distributions of the measurement results (red dots in \fref{fig:probability exp}(a)) at different $x\in[-\frac{\pi}{2},0]$(it is symmetrical for $x\in [0,\frac{\pi}{2}]$). For comparison, we also carry out the experiments with the same control 
at some non-sweet spot in time, and plot the probabilities of the measurements results as black dots in \fref{fig:probability exp}(a). It can be seen that at the non-sweet spot in time the periods of the distributions get larger and the interference visibility gets smaller when the actual value of $x$ deviates from $0$. However, at the sweet spots in time the probability fringes remain the same for all values of $x$. In \fref{fig:probability exp}(b), we plot the QFI for the case of $N=8$, it can be seen that at the sweet spot in time $T_8=4\pi$, $\sqrt{J}$ is close to $\sqrt{16N^2}= 32$ for all x(here $N=8$), while at the non-sweet spot in time, only when $x$ is near $0$, $\sqrt{J}$ is close to the optimal value $\sqrt{16N^2\sin^2\frac{4\pi}{8}}\approx 22.63$, the value decreases when $x$ deviates from $0$.

It is worth to mention that in the control-enhanced sequential scheme the optimal measurements are simple local projective measurements, which can be easily implemented with high quality(see Supplemental material). For example, for the case of $N=8$ the visibility in our experiment is larger than $0.984$, while the visibility of the post-selected $N-$photon entangled states decreases rapidly when $N$ increases\cite{Wang16tenphoton}.  

\emph{Discussion}--We provided an optimal procedure for a scalable control-enhanced sequential scheme that can achieve the Heisenberg precision for general dynamics.
We experimentally implemented the scheme for the estimation of the orientation of a phase plate, and showed that the scheme can achieve the Heisenberg precision for general noncommuting dynamics. We also identified the sweet spots in time at which the scheme can be realized with pre-fixed controls without any adaptation. This pushes forward both theoretical and experimental studies of quantum metrology under general non-commuting dynamics. We expect the results will have wide implications in various applications of quantum metrology.


%

The work at USTC is supported by the National Key Research And Development Program of China (Grant No.2018YFA0306400), the National Natural Science Foundation of China under Grants (Nos. 11574291, 11774334, 61327901 and 11774335),  the National Key Research and Development Program of China (No.2017YFA0304100),  Key Research Program of Frontier Sciences, CAS (No.QYZDY-SSW-SLH003), Anhui Initiative in Quantum Information Technologies and  China Postdoctoral Science Foundation (Grant Nos.2016M602012 and 2018T110618). The work at CUHK is supported by the Research Grants Council of Hong Kong(GRF No. 14207717).

 \clearpage
 \newpage
 \addtolength{\textwidth}{-1in}
 \addtolength{\oddsidemargin}{0.5in}
 \addtolength{\evensidemargin}{0.5in}

 \setcounter{equation}{0}
 \setcounter{figure}{0}
 \setcounter{table}{0}
 \setcounter{section}{0}
 \makeatletter
 \renewcommand{\theequation}{S\arabic{equation}}
 \renewcommand{\thefigure}{S\arabic{figure}}
 \renewcommand{\thetable}{S\arabic{table}}
 \renewcommand{\thesection}{S\arabic{section}}
\onecolumngrid
 \begin{center}
 	\textbf{\large Control-enhanced sequential scheme for general quantum parameter estimation at the Heisenberg limit: Supplementary Materials}
 \end{center}

\section{S1. Bloch vectors of generators for time-independent Hamiltonian}

Here we consider a general Hamiltonian $H$ on a qubit, which can be written as $H=\vec{h}\cdot\vec{\sigma}$ with $\vec{\sigma}=(\sigma_1,\sigma_2,\sigma_3),$ where
\[\sigma_1=\left(
\begin{array}{cc}
0 & 1 \\
1 & 0 \\
\end{array}
\right),\quad
\sigma_2=\left(
\begin{array}{cc}
0 & -\rmi \\
\rmi & 0 \\
\end{array}
\right),\quad
\sigma_3=\left(
\begin{array}{cc}
1 & 0 \\
0 & -1 \\
\end{array}
\right)\] are Pauli matrices. Then we have $V_0=\partial_{\phi}H=\vec{v_0}\cdot\vec{\sigma}$ with $\vec{v_0}=\partial_\phi\vec{h}$, $V_{t}=U^\dagger_{t}(\phi) V_0 U_{t}(\phi)=\vec{v}_t\cdot\vec{\sigma}$ with $\vec{v}_t=R_{\vec{n}_h}(wt)\vec{v}_0$, where $R_{\vec{n}_h}(t)$ is the rotation on the Bloch vector generated by the unitary $U_{t}(\phi)$, with the rotating axis as $\vec{n}_h=\vec{h}/|\vec{h}|$(a unit vector along $\vec{h}$) and the rotating angular speed as $w=2|\vec{h}|$. $\vec{v}_t$ can be decomposed into two components as $\vec{v}_t=\vec{v}_{t,\parallel}+\vec{v}_{t,\perp}$, where the component parallel to the rotating axis $\vec{n}_h$ does not change, i.e., $\vec{v}_{t,\parallel}=\vec{v}_{0,\parallel}=\vec{v}_0\cdot\vec{n}_h$, the component perpendicular to the axis rotates with an angular speed $w$, $\vec{v}_{t,\perp}=|\vec{v}_{0,\perp}|(\cos wt\vec{n}_1^\perp+\sin wt\vec{n}_2^\perp)$, here $\vec{v}_{0,\perp}=\vec{v}_0-(\vec{v}_0\cdot\vec{n}_{h})\vec{n}_{h}, \vec{n}_1^\perp=\vec{v}_{0,\perp}/|\vec{v}_{0,\perp}|, \vec{n}_2^\perp=\vec{n}_h\times \vec{n}_1^\perp$. Accordingly, the accumulated generator has a parallel and a perpendicular components, i.e.,  $\vec{s}_T=\vec{s}_{T,\parallel}+\vec{s}_{T,\perp}$, where $\vec{s}_{T,\parallel}=\int_{0}^{T}\vec{v}_{t,\parallel} d{t}=\vec{v}_{0,\parallel} T$ is a uniform rectilinear motion along $\vec{n}_h$ and  $\vec{s}_{T,\perp}=\int_{0}^{T}\vec{v}_{t,\perp} d{t}=|\vec{v}_{0,\perp}|/w[\sin wt\vec{n}_1^\perp+(1-\cos wt)\vec{n}_2^\perp]$ is a circular motion on the plane spanned by $\vec{n}_1^\perp$ and $\vec{n}_2^\perp.$ Hence, $\vec{s}_T$ is a helical line. The variance of $S_T$ is $\<\Delta S_T^2\>=|\vec{s}_T|^2-\bra{\psi}S_T\ket{\psi}^2$, which is upper bounded by $|\vec{s}_T|^2=\left|\vec{v}_{0,\parallel} \right|^2T^2+4\frac{\left|\vec{v}_{0,\perp}\right|^2}{w^2}\sin^2 \frac{wT}{2}$. This is always smaller than $|\left|\vec{v}_0\right|^2T^2$, which is the maximal that one can achieve.


\section{S2. Optimal probe states and optimal measurements}
In this section, we provide a geometrical method for identifying the optimal probe states and optimal measurements.

We first write the Hamiltonian, $H(x)=\sin2x\sigma_1+\cos2x\sigma_3$, as $H(x)=\hat{\vec{n}}_h(x)\cdot\vec{\sigma}$ with $\hat{\vec{n}}_h(x)=(\sin 2x, 0, \cos 2x)$. Let $\hat{\vec{n}}_{1}(x)=(\cos 2x, 0, -\sin 2x) $, $\hat{\vec{n}}_{2}=\hat{\vec{n}}_h\times\hat{\vec{n}}_{1}=(0, 1, 0)$, $\{\hat{\vec{n}}_h(x),\hat{\vec{n}}_{1}(x),\hat{\vec{n}}_{2}\}$ then form a basis on the Bloch sphere.

Under the dynamics, $U_t(x)=e^{-\rmi\hat{\vec{n}}_h(x)\cdot\vec{\sigma}t}$, the output state is $\rho_{t}(x)=U_t(x)\rho_{0}U_t^\dagger(x)$, here the initial probe state $\rho_{0}$ can be written as $\rho_0=\frac{1}{2}(1+\hat{\vec{n}}_{in}\cdot\vec{\sigma})$. The optimal probe that leads to the maximal QFI is the state that maximizes the Bures distance between $\rho_{T}(x)$ and its neighboring state $\rho_{t}(x+dx)=U_t(x+dx)\rho_{0}U_t^\dagger(x+dx)$. This is equivalent to maximize the distance between $\rho_{0}$ and $[U_t^\dagger(x)U_t(x+dx)]\rho_{0}[U_t^\dagger(x)U_t(x+dx)]^\dagger$. Let $U'=U_t^\dagger(x)U_t(x+dx)=e^{-\rmi t^\prime\hat{\vec{n}}_{U^\prime}\cdot\vec{\sigma}}$, where $t^\prime=2\sin t dx$ and $$\hat{\vec{n}}_{U^\prime}=\cos t\hat{\vec{n}}_{1}(x)-\sin t \hat{\vec{n}}_{2}.$$
The optimal probe states can then be easily identified as the pure states with the unit Bloch vector perpendicular to $\hat{\vec{n}}_{U^\prime}$ (as they have the largest change under $e^{-\rmi t^\prime\hat{\vec{n}}_{U^\prime}\cdot\vec{\sigma}}$), which can be written as $\rho_{0}=\frac{1}{2}(1+\hat{\vec{n}}_{in}\cdot\vec{\sigma})$ with
	\begin{equation}\label{eq:optimal probe states}
	\hat{\vec{n}}_{in}=\cos\alpha\hat{\vec{n}}_h(x)+\sin\alpha\hat{\vec{n}}_{3},
	\end{equation}
	here $\hat{\vec{n}}_{3}=\hat{\vec{n}}_h\times\hat{\vec{n}}_{U^\prime}=\sin t\hat{\vec{n}}_{1}(x)+\cos t \hat{\vec{n}}_{2}$, $\alpha$ can be chosen arbitrarily.
The optimal measurement in this case is the projective measurement along the direction which is also perpendicular to $\hat{\vec{n}}_{U^\prime}$,
	\begin{equation}\label{eq:optimal measurement}
\hat{\vec{n}}_M=\cos\beta\hat{\vec{n}}_h(x)+\sin\beta\hat{\vec{n}}_{3}
	\end{equation}
 where $\beta$ can be chosen arbitrarily.
 \begin{figure}[t]
 	\center{\includegraphics[scale=0.6]{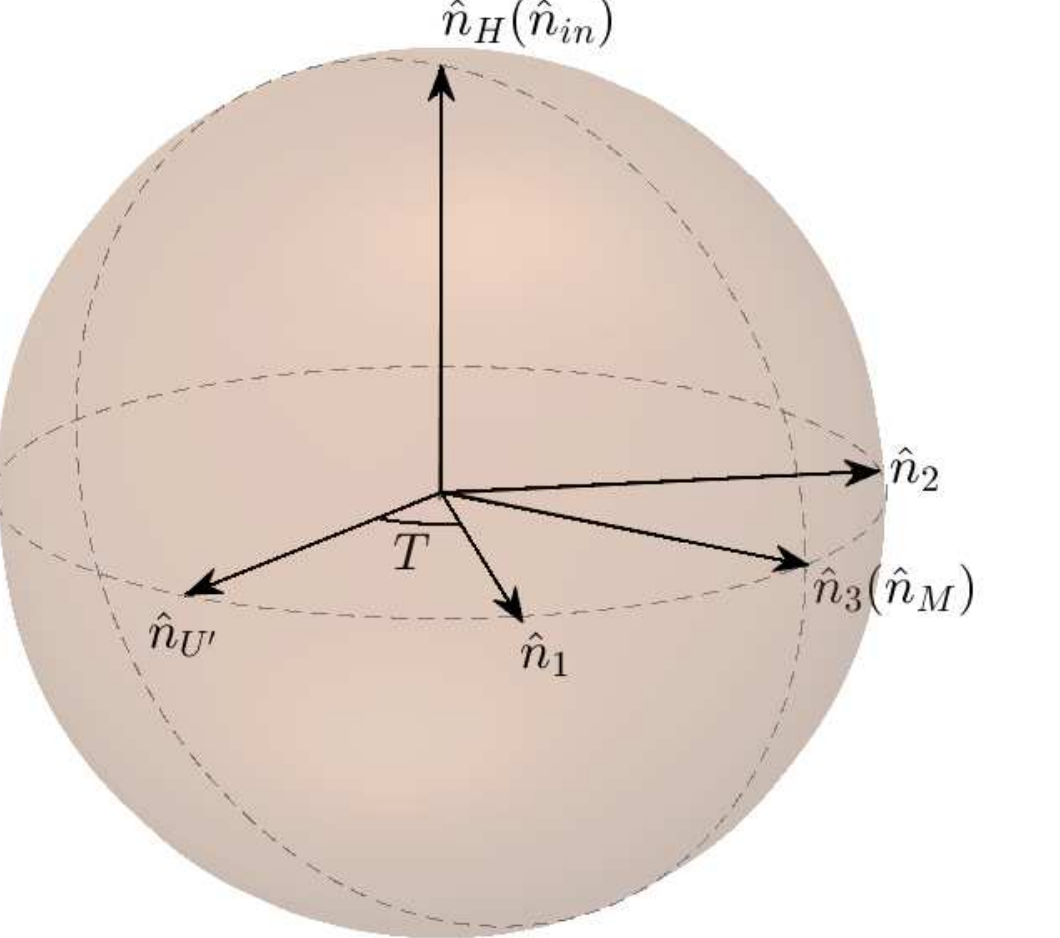}}
 	\caption{\label{fig:geometry}{Geometrical picture of optimal probe states and measurements.}}
 \end{figure}

Under the control-enhanced scheme, the probe state passes through the dynamics, $U_t(x)$ and the control, $U_c$, $N$ times. If $U_c$ makes the dynamics commute, the precision limit is determined by
\begin{eqnarray}
\aligned
\{[U_cU_t(x)]^N\}^\dagger [U_cU_t(x+dx)]^N=&\{[U_cU_t(x)]^\dagger [U_cU_t(x+dx)]\}^N\\
=&[U_t^\dagger(x)U_t(x+dx)]^N\\
=&U'^N=e^{-\rmi Nt^\prime\hat{\vec{n}}_{U^\prime}\cdot\vec{\sigma}}.
\endaligned
\end{eqnarray}
$\hat{\vec{n}}_{U^\prime}$ remains as the same, the optimal states and optimal measurements are thus also the same.

In the experiment, $\alpha$ is taken as $0$, thus $\hat{n}_{in}=\hat{n}_h(x)=(\sin 2x, 0, \cos 2x)$. $\beta$ is mostly taken as $\pi/2$, $\hat{n}_M=\sin t\hat{n}_1(x)+\cos t\hat{n}_2$. This corresponds to the optimal probe state taken as $\cos x|0\rangle+\sin x|1\rangle$ and the projective measurements along the direction $\hat{n}_M=(\sin t\cos 2x, \cos t, -\sin t \sin2x)$. They in general depend on the value of the parameter, thus can only be prepared adaptively. However, at some specific time points, the optimal probe state and measurement can be pre-fixed without the need of adaptation. Specifically, when $t=\pi/2$, $\hat{\vec{n}}_{U^\prime}=\cos t\hat{\vec{n}}_{1}(x)-\sin t \hat{\vec{n}}_{2}=-\hat{\vec{n}}_{2}$, which is independent of $x$. In this case the optimal probe state can be any state with $\hat{\vec{n}}_{in}=(\sin\alpha, 0, \cos {\alpha})$, and the optimal measurement as any projective measurement along the direction $\hat{\vec{n}}_M=(\sin \beta, 0, \cos\beta)$, where $\alpha$ and $\beta$ are independent of the actual value of $x$ and can be chosen arbitrary. In the experiment, we take $\alpha=0$ and $\beta=\frac{\pi}{2}$. At the sweet spot in time the optimal probe state is then taken as $|H\rangle$ and the optimal measurement is the projective measurement along the direction of $\sigma_1$.

\begin{figure*} [t]
	\center{\includegraphics[scale=0.5]{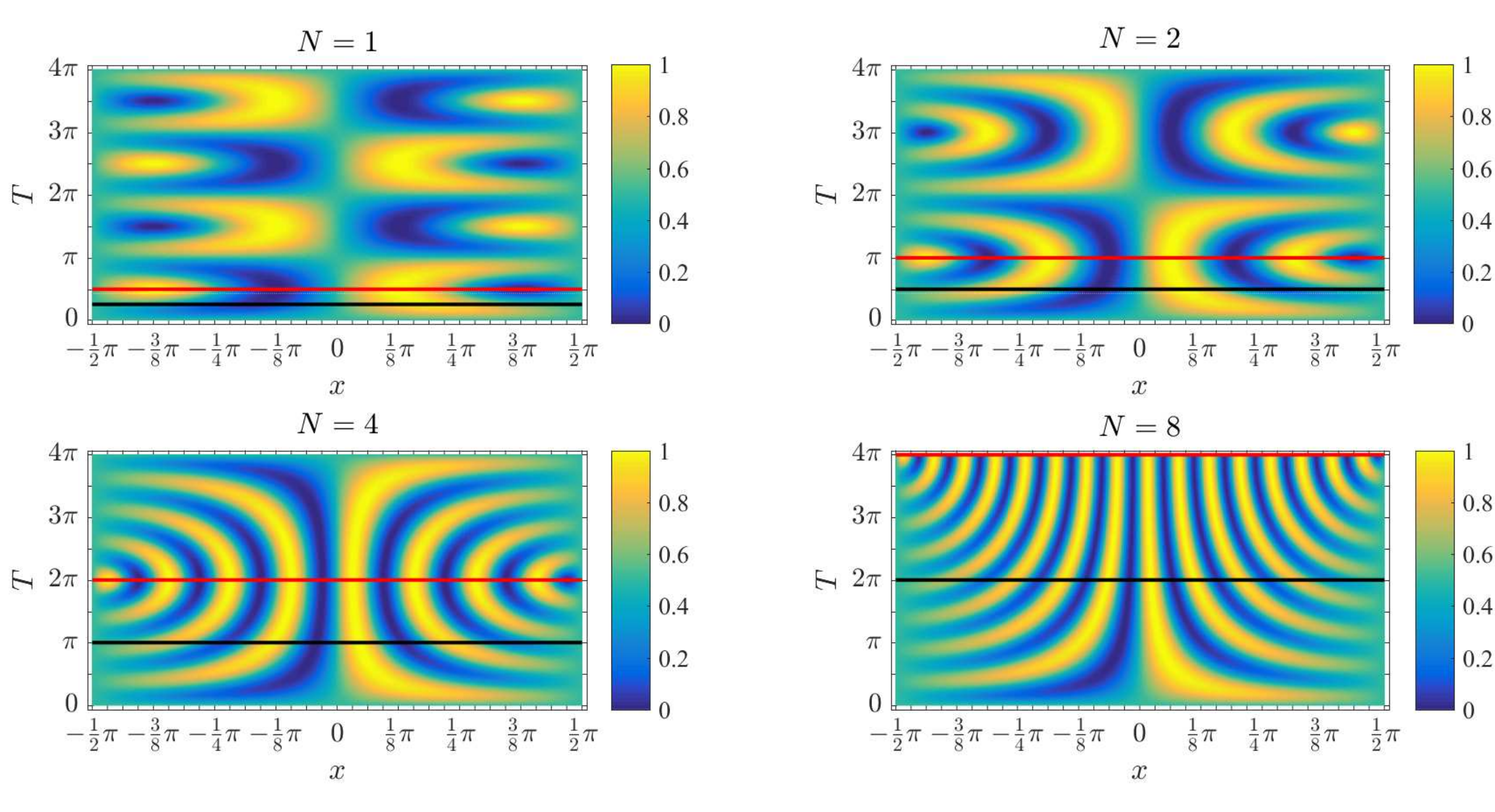}}
	\caption{\label{fig:probability 3d}(color online)Theoretical Probability distribution with respect to $x$ and $T$.}
\end{figure*}

\section{S3. Adaptive controls and sweet spots in time}
In the control-enhanced sequential scheme, the optimal probe state, control and measurement typically all need to be implemented adaptively based on the estimation, $\hat{x}$, obtained from previously measured data. We show how the difference between $\hat{x}$ and $x$ affects the performance.

When the probe state goes through $N$ passes of the dynamics ($U_t(x)$) and the control ($U_c(\hat{x})$), the output state is given by
\begin{equation}\label{eq:output state}
\rho_{out}(x)=[U_c(\hat{x})U_t(x)]^N\rho_{0}[U_t^\dagger(x) U_c^\dagger(\hat{x})]^N.
\end{equation}
here $\rho_0=\frac{1}{2}(1+\hat{\vec{n}}_{in}(\hat{x})\cdot\vec{\sigma})$, with $\hat{\vec{n}}_{in}(\hat{x})$ optimally chosen based on the estimation $\hat{x}$, the control is taken as $U_c(\hat{x})=U_t^\dagger(\hat{x})=e^{\rmi\hat{\vec{n}}_h(\hat{x})\cdot\vec{\sigma} t}$ with $\hat{\vec{n}}_h(\hat{x})=(\sin 2\hat{x}, 0, \cos 2\hat{x})$. Thus
\begin{eqnarray}
\aligned
U_c(\hat{x})U_t(x)&=U_t^\dagger(\hat{x})U_t(x)\\
&=e^{iH(\hat{x})t}e^{-iH(x)t}\\
&=e^{-it_e\hat{\vec{n}}_e\cdot \vec{\sigma}},
\endaligned
\end{eqnarray}
here $t_e$ and $\hat{\vec{n}}_e$ are determined from the following equations
\begin{eqnarray}
\aligned
  \cos t_e&=\cos^2t+\hat{\vec{n}}_h(\hat{x})\cdot\hat{\vec{n}}_h(x)\sin^2t=\cos^2t+\cos2(x-\hat{x})\sin^2t,\\
  \sin t_e\hat{\vec{n}}_e&=\sin t\cos t[\hat{\vec{n}}_h(x)-\hat{\vec{n}}_h(\hat{x})]+\sin^2 t\hat{\vec{n}}_h(x)\times\hat{\vec{n}}_h(\hat{x}).
\endaligned
\end{eqnarray}

By performing the projective measurement along the direction $\hat{\vec{n}}_M(\hat{x})\cdot\vec{\sigma}$ on the output state (here $\hat{\vec{n}}_M(\hat{x})$ is optimally chosen based on $\hat{x}$ as in Eq.(\ref{eq:optimal measurement})), one gets two outcomes with the probabilities $P_+$ and $P_-=1-P_+$, where
\begin{align}\label{eq: probability}
P_+=0.5+\frac{A_N^2}{8}\sin 2\alpha\sin^22t[\cos 2(x-\hat{x})-1]^2+A_N\sin t\sin 2(x-\hat{x})\cos Nt_e,
\end{align}
here $A_N\equiv\frac{\sin Nt_e}{\sin t_e}=\sum_{k=0}^{N-1}\cos(N-1-k)t_e\cos^kt_e$, $\alpha$ is the same as in \eref{eq:optimal probe states} which can be chosen arbitrarily. In the experiment, it is chosen as 0. 

As $P_+$ is only determined by the difference between $x$ and $\hat{x}$, without loss of generality, we can set $\hat{x}=0$, then
$P_+=0.5+ A_N\sin t\sin 2x\cos Nt_e$. The Fisher information can then be calculated as $F=(\frac{\partial P_+}{\partial x})^2/[P_+(1-P_+)]$. We plot the probability distribution and the Fisher information with respect to $x$ and $T=Nt$ in \fref{fig:probability 3d} (note that the value of $x$ represents the difference between $\hat{x}$ and $x$ since we have set $\hat{x}=0$).

\begin{figure*}[t]
	\center{\includegraphics[scale=0.5]{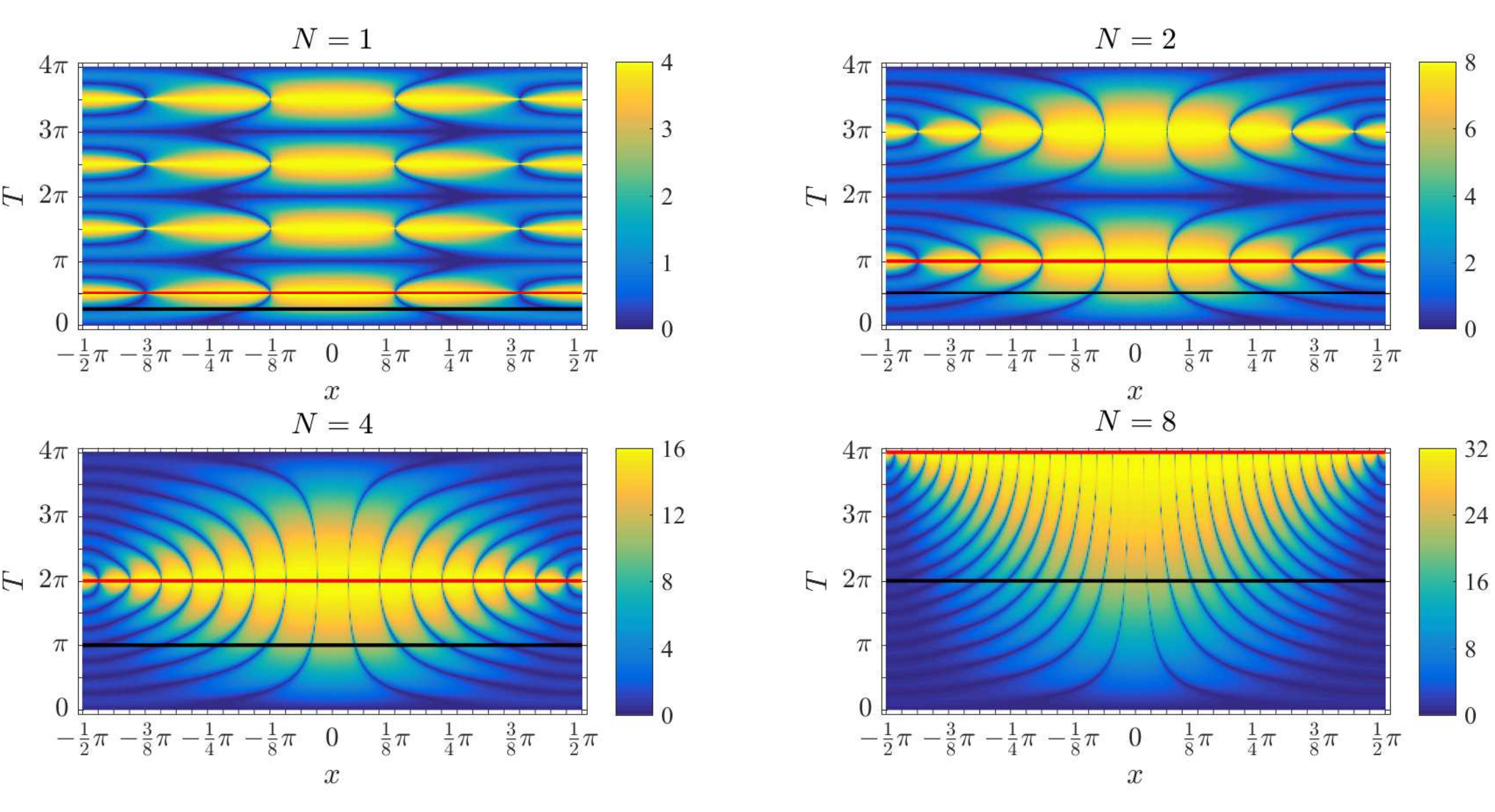}}
	\caption{\label{fig:Fisher 3d}(color online)Theoretical Fisher Information with respect to $x$ and $T$.}
\end{figure*}

From \fref{fig:probability 3d}, it can be seen that the Fisher information only gets much smaller than the optimal value (which achieves at $x=0$) when $x$ is significantly away from $0$, showing that the adaptive controls are quite robust. In particular at the sweet spots in time($t=\frac{\pi}{2}N$), the Fisher information does not change with $x$. This can also be seen by a direct calculation. At the sweet spot in time, $t=\frac{\pi}{2}N$, $t=\frac{t}{N}=\pi/2$, $\cos t_e=\cos 2x$ ($\hat{x}$ has been set as 0), $P_+=0.5+ 0.5\sin 4Nx$, thus $F=(\frac{\partial P_+}{\partial x})^2/[P_+(1-P_+)]=16N^2$, which is independent of the actual value of $x$.
This is also clearly shown by the fringes at $t$ in \fref{fig:probability 3d}, where the periods remain the smallest and the interference visibility is 100\% for all $x$, showing the control designed according to $\hat{x}=0$ is optimal for all $x$ at $t$. This can also be clearly seen in \fref{fig:Fisher 3d} where the Fisher information achieves the largest value for all $x$ at $t$.

Moreover, at these time points, the optimal probe states and optimal measurements can also be pre-fixed. Specifically, at these time points the optimal probe state can all be taken as $|\psi\rangle=|0\rangle$ and the optimal measurement as the projective measurement on the eigenvectors of $\sigma_1$, which are independent of the actual value of the parameter. Thus under the control-enhanced sequential scheme with $N$ passes, not only the QFI achieves the maximal value at the time point $t=\frac{\pi}{2}N$, but the optimal probe state, optimal control and optimal measurement can all be pre-fixed without the need of adaptation. These time points are thus referred as the sweet spots in time.

\section{S4. Experimental implementation of the optimal probe states, controls and measurements}

For the control-enhanced sequential scheme, the optimal probe states, controls and measurements typically depend on $x$ and need to be updated adaptively based on the estimation, $\hat{x}$, obtained from previously measured data. As shown in \fref{fig:setup}, to adaptively prepare the optimal probe state, which is $\ket{\psi}=\cos \hat{x}\ket{0}+\sin \hat{x}\ket{1}$, the rotation angle of HWP1 is set as $\frac{\hat{x}}{2}$ and the rotation angle of QWP1 is set as $\hat{x}$\cite{hou2016error}. To adaptively implement the optimal control, which is $U_c=U^\dagger_t(\hat{x})=e^{\rmi (\sin2\hat{x}\sigma_1+\cos2\hat{x}\sigma_3)t}$, the rotation angles of QWP2 and QWP3 are set as $\hat{x}+\frac{\pi}{4}$ and the rotation angle of HWP2 is set as $\hat{x}-\frac{t}{2}-\frac{\pi}{4}$ \cite{simon1990minimal}. To perform the optimal measurement, which is the projective measurement on the eigenvectors of $\sin \frac{T}{N}\cos 2\hat{x} \sigma_1+\cos  \frac{T}{N}\sigma_2-\sin  \frac{T}{N}\sin 2\hat{x}\sigma_3$, the rotation angle of HWP3 is set as $\frac{\hat{x}}{2}-\frac{t}{4}+\frac{\pi}{4}$ and the rotation angle of QWP4 is set as $\hat{x}+\frac{\pi}{4}$\cite{hou2016error}.

An automatic control system is designed to update the probe state, control and measurement adaptively. The control system includes a coincidence unit, an executive Labview program and seven motorized stages. The coincidence unit heralds the generation of the probe photons and collects the measurement results. The executive Labview program then analyzes the measurement data collected by the coincidence unit, obtains the estimation of the parameter ($\hat{x}$), and updates the rotation angles of the wave-plates. The information is then sent to the rotation stages to rotate the three HWPs and four QWPs to the desired angles. With this design, the adaptive procedure can be automatically realized. 

\section{S5. Error analysis}

In the experiment, we get one estimation of the parameter, $\hat{x}$, based on $50$ measurement outcomes. To experimentally obtain the standard deviation of this estimation, we repeat this $50$ measurements $1000$ times and obtain $1000$ realizations of $\hat{x}$ to get the distribution of $\hat{x}$. The standard deviation, $\delta\hat{x}$ is then obtained from the distribution. The error for this experimentally obtained $\delta\hat{x}$, $\Delta(\delta\hat{x})$, is well approximated by $\Delta(\delta\hat{x})=\frac{\delta\hat{x}}{\sqrt{2(K-1)}}$ with $K=1000$ \cite{ahn2003standard}, which is used for draw the error bar in Fig{\ref{fig:adaptive}. The experimental Fisher information is obtained directly from the $\delta\hat{x}$ as $\sqrt{J}=\frac{1}{\delta\hat{x}\sqrt{n}}$ with $n=50$. The error for $\sqrt{J}$ is well approximated by $\Delta{(\sqrt{J})}=\frac{\sqrt{J}}{\sqrt{2(K-1)}}$.

\section{S6. Number of controls using four-mirror cavity loop}

Here we show how to deterministically control the number of passes through the cavity loop without affecting the positions and the directions of the input and output beams. With this design, the change of the number of passes do not affect the modules of the preparation and the measurement, which eases the implementation of the experiment.

We use four mirrors to control the number of passes. The four mirrors, denoted as M$_1$ to M$_4$ are laid out as in \fref{fig:cavity}(a). The positions of M$_1$, M$_3$ and M$_4$ are fixed in the experiments. The number of loops are controlled by translationally moving M$_2$, which is installed on a translation stage.

The largest $N$ realizable is limited by the size of the clear apertures of optical devices in the cavity and the diameter of the beam. In our experiment, the clear apertures of the wave-plates have a diameter of 20 mm. The largest diameter of the photon beams, which happens for $N=8$, is not larger than 1.6 mm.

The position of M$_2$ controls the distance between adjacent beams, which determines the number of loops. We denote the length of DF in Fig\ref{fig:cavity} as $d_4$, the length of BE as $d_2$. The distance between two adjacent beams(denoted as $d$) is determined by $d_4$ and $d_2$ as
\begin{equation}\label{eq:cavity}
d=d_4-d_2.
\end{equation}
And for $N$ loops, we have $d_4=Nd$.

We note that the distance between the adjacent beams need to be larger than the diameter of the photon beams. Thus in the experiment, we choose $d_4=16$ mm, which corresponds to $d=2>1.6$ mm for $N=8$. From Eq.(\ref{eq:cavity}), it is easy to see that to realize $N$ loops, M$_2$ just needs to be moved translationally with $d_2=(1-\frac{1}{N})d_4$. Thus to realize $N=1, 2,4,8$ loops, we just move M$_2$ with $d_2=0, 8, 12, 14$ mm, respectively.



\begin{figure*}[t]
	\center{\includegraphics[scale=0.5]{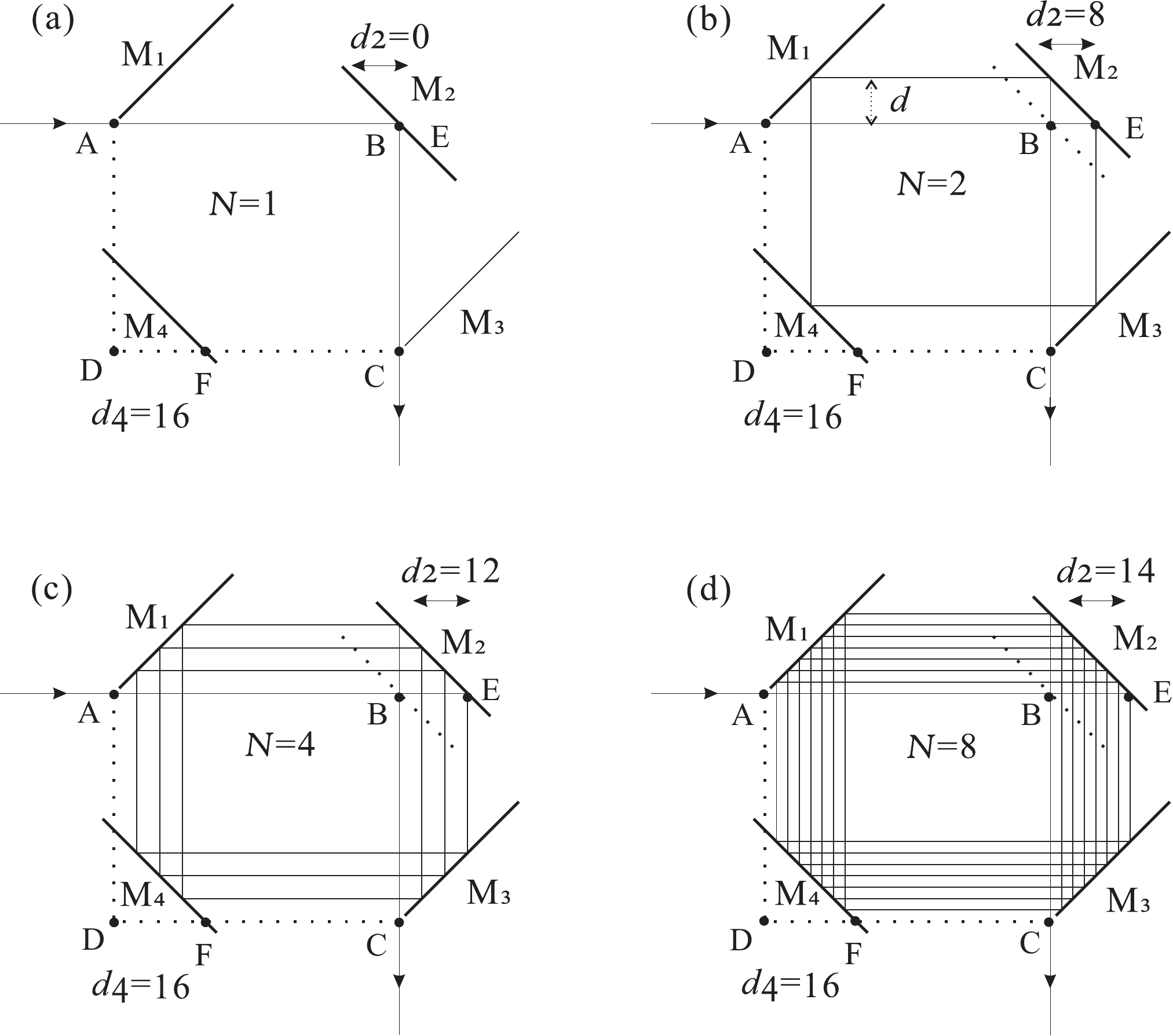}}
	\caption{\label{fig:cavity}{Control the number of pass through a four-mirror cavity }}
\end{figure*}

\section{S7. Shot-noise limit}
In the main context we presented the precision limit for the control-enhanced scheme for a non-commuting dynamics
\begin{equation}\label{eq:unitary operator}
U_t(x)=e^{-it(\sin2x\sigma_1+\cos2x\sigma_3)},
\end{equation}
and compared it with the shot-noise limit. Here we show how the shot-noise limit is obtained.

The shot-noise limit is obtained by dividing the total time $T$ into $N$ slices with each slice acting on a probe state independently. The maximal QFI for each slice is $J_t=16\sin^2t$, here $t=\frac{T}{N}$. The QFI for $N$ independent repetition is simply $N$ times $J_t$, which is
$J=NJ_t=16N\sin^2\frac{T}{N}.$
We associate the shot-noise limit with the maximal QFI that can be achieved by maximizing over $N$,
\begin{equation}\label{eq:shot-noise limit for T}
J_{shot}=\max_N 16N\sin^2\frac{T}{N}.
\end{equation}
This maximum value is achieved when $\frac{\partial J}{\partial N}=16\sin t(\sin t-2t\cos t)=0$, where $t=\frac{T}{N}$. This gives the optimal value of $t$ as $t_0\approx 1.1656$, which leads to the maximal value of $16N\sin^2\frac{T}{N}$ as $\sim 11.593T$. This provides an upper bound on $J_{shot}$. Since $N$ is an integer, $J_{shot}$ is actually $\max(J(N_-,T),J(N_+,T))$ with $N_-=\max(1,\lfloor T/t_0\rfloor)$ and $N_+=\max(1,\lceil T/t_0\rceil)$, the numerical simulation in \fref{fig:J_stand}  shows that this is only slightly smaller than $11.593T$ and converges to $11.593T$ when $T$ increases. 

\begin{figure*}[t]
	\center{\includegraphics[scale=0.7]{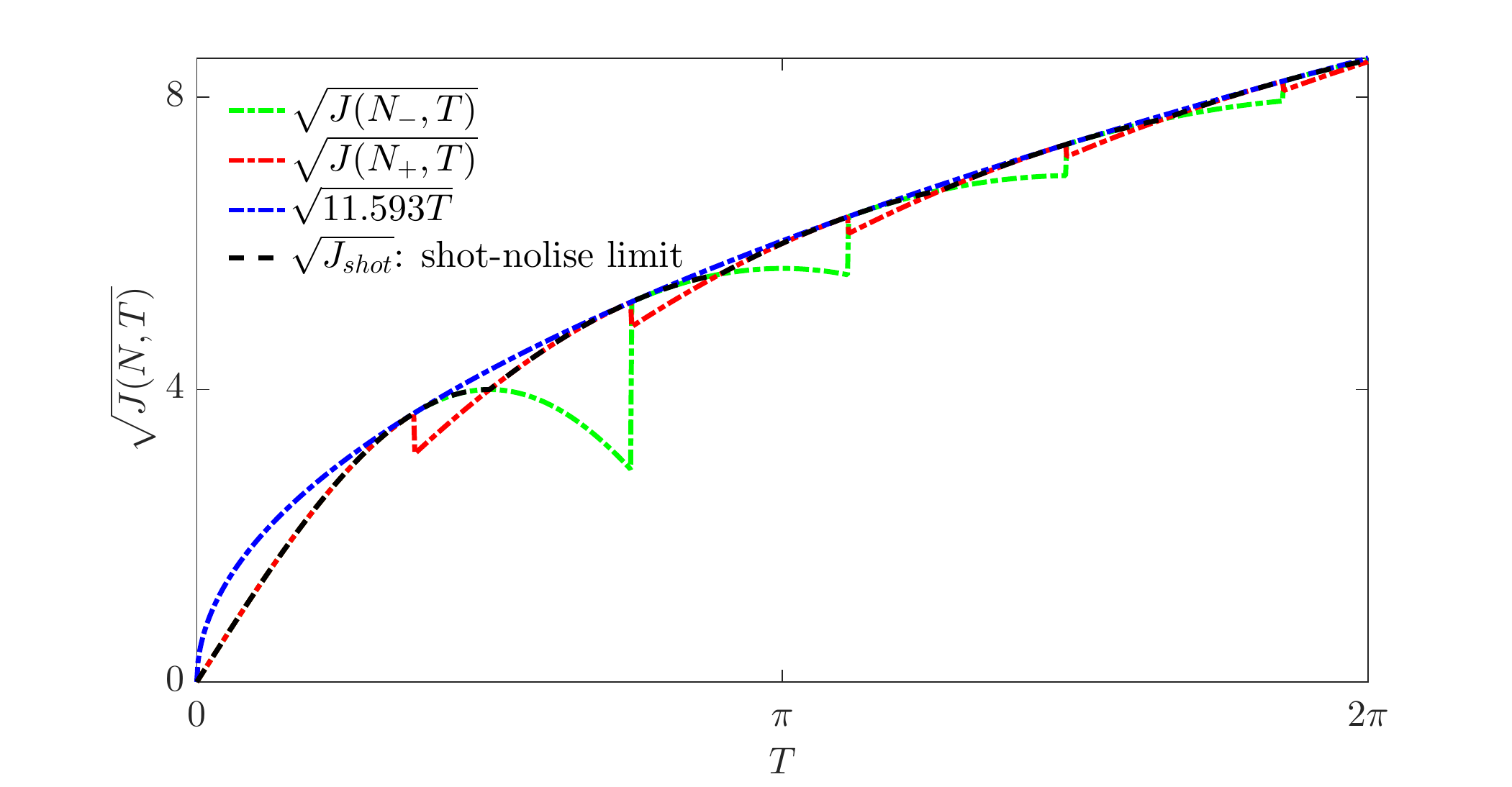}}
	\caption{\label{fig:J_stand}(color online) {The shot-noise limit of the non-commuting dynamics. }}
\end{figure*}

\section{S8. Prior information}
For the sequential scheme with adaptive control, the parameter has some prior distribution which also contributes to the precision. Denote $p(x)$ as the prior distribution, if it has finite support within certain interval, for example $[0,\frac{\pi}{8}]$, then the variance $E[(\hat{x}-x)^2]=\int_0^{\frac{\pi}{8}} p(x)(\hat{x}-x)^2 dx$ is already finite even for a guess $\hat{x}=\frac{\pi}{16}$ without any measurement data. In general with a prior distribution,
\begin{equation}
\label{eq:prior}
\delta \hat{x}\geq \frac{1}{\sqrt{n\int p(x)J(x)dx+F_p(x)}},
\end{equation}
where $n$ is the number of times the measurement is repeated, $J(x)$ is the quantum Fisher information of the output state $\rho(x)$, and $F_p(x)=\int p(x)(\frac{\partial ln p(x)}{\partial x})^2 dx$ is the classical Fisher information of the prior distribution \cite{Tsang11fundamental}.
When $n$ gets large, the quantum Fisher information typically dominates, Eq(\ref{eq:prior}) reduces to the usual quantum Cramer-Rao bound. 
However, there are some special point at which $J(x)\rightarrow 0$, for example, in the case of sequential scheme with one control, the quantum Fisher information is 0 when $T=k\pi$. At these points, the contribution of the classical Fisher information can not be neglected even when $n$ is large. This can be seen in \fref{fig:adaptive} where the experimental precision exceeds the quantum Fisher information near $T=0, \pi$ and $2\pi$. The contribution of the prior distribution is negligible at other points where $J(x)$ is not close to $0$.

%



\begin{thebibliography}{30}
	\expandafter\ifx\csname natexlab\endcsname\relax\def\natexlab#1{#1}\fi
	\expandafter\ifx\csname bibnamefont\endcsname\relax
	\def\bibnamefont#1{#1}\fi
	\expandafter\ifx\csname bibfnamefont\endcsname\relax
	\def\bibfnamefont#1{#1}\fi
	\expandafter\ifx\csname citenamefont\endcsname\relax
	\def\citenamefont#1{#1}\fi
	\expandafter\ifx\csname url\endcsname\relax
	\def\url#1{\texttt{#1}}\fi
	\expandafter\ifx\csname urlprefix\endcsname\relax\def\urlprefix{URL }\fi
	\providecommand{\bibinfo}[2]{#2}
	\providecommand{\eprint}[2][]{\url{#2}}
	
	\bibitem[{\citenamefont{Giovannetti et~al.}(2004)\citenamefont{Giovannetti,
			Lloyd, and Maccone}}]{GiovLM04}
	\bibinfo{author}{\bibfnamefont{V.}~\bibnamefont{Giovannetti}},
	\bibinfo{author}{\bibfnamefont{S.}~\bibnamefont{Lloyd}}, \bibnamefont{and}
	\bibinfo{author}{\bibfnamefont{L.}~\bibnamefont{Maccone}},
	\bibinfo{journal}{Science} \textbf{\bibinfo{volume}{306}},
	\bibinfo{pages}{1330} (\bibinfo{year}{2004}).
	
	\bibitem[{\citenamefont{Nagata et~al.}(2007)\citenamefont{Nagata, Okamoto,
			O'brien, Sasaki, and Takeuchi}}]{Naga07beating}
	\bibinfo{author}{\bibfnamefont{T.}~\bibnamefont{Nagata}},
	\bibinfo{author}{\bibfnamefont{R.}~\bibnamefont{Okamoto}},
	\bibinfo{author}{\bibfnamefont{J.~L.} \bibnamefont{O'brien}},
	\bibinfo{author}{\bibfnamefont{K.}~\bibnamefont{Sasaki}}, \bibnamefont{and}
	\bibinfo{author}{\bibfnamefont{S.}~\bibnamefont{Takeuchi}},
	\bibinfo{journal}{Science} \textbf{\bibinfo{volume}{316}},
	\bibinfo{pages}{726} (\bibinfo{year}{2007}).
	
	\bibitem[{\citenamefont{Higgins et~al.}(2007)\citenamefont{Higgins, Berry,
			Bartlett, Wiseman, and Pryde}}]{HiggBBW07}
	\bibinfo{author}{\bibfnamefont{B.~L.} \bibnamefont{Higgins}},
	\bibinfo{author}{\bibfnamefont{D.~W.} \bibnamefont{Berry}},
	\bibinfo{author}{\bibfnamefont{S.~D.} \bibnamefont{Bartlett}},
	\bibinfo{author}{\bibfnamefont{H.~M.} \bibnamefont{Wiseman}},
	\bibnamefont{and} \bibinfo{author}{\bibfnamefont{G.~J.} \bibnamefont{Pryde}},
	\bibinfo{journal}{Nature} \textbf{\bibinfo{volume}{450}},
	\bibinfo{pages}{393} (\bibinfo{year}{2007}).
	
	\bibitem[{\citenamefont{Giovannetti et~al.}(2011)\citenamefont{Giovannetti,
			Lloyd, and Maccone}}]{GiovLM11}
	\bibinfo{author}{\bibfnamefont{V.}~\bibnamefont{Giovannetti}},
	\bibinfo{author}{\bibfnamefont{S.}~\bibnamefont{Lloyd}}, \bibnamefont{and}
	\bibinfo{author}{\bibfnamefont{L.}~\bibnamefont{Maccone}},
	\bibinfo{journal}{Nature Photonics} \textbf{\bibinfo{volume}{5}},
	\bibinfo{pages}{222} (\bibinfo{year}{2011}).
	
	\bibitem[{\citenamefont{Xiang et~al.}(2011)\citenamefont{Xiang, Higgins, Berry,
			Wiseman, and Pryde}}]{XianHBW11}
	\bibinfo{author}{\bibfnamefont{G.~Y.} \bibnamefont{Xiang}},
	\bibinfo{author}{\bibfnamefont{B.~L.} \bibnamefont{Higgins}},
	\bibinfo{author}{\bibfnamefont{D.~W.} \bibnamefont{Berry}},
	\bibinfo{author}{\bibfnamefont{H.~M.} \bibnamefont{Wiseman}},
	\bibnamefont{and} \bibinfo{author}{\bibfnamefont{G.~J.} \bibnamefont{Pryde}},
	\bibinfo{journal}{Nat. Photonics} \textbf{\bibinfo{volume}{5}},
	\bibinfo{pages}{43} (\bibinfo{year}{2011}).
	
	\bibitem[{\citenamefont{Slussarenko et~al.}(2017)\citenamefont{Slussarenko,
			Weston, Chrzanowski, Shalm, Verma, Nam, and Pryde}}]{Slus17unconditional}
	\bibinfo{author}{\bibfnamefont{S.}~\bibnamefont{Slussarenko}},
	\bibinfo{author}{\bibfnamefont{M.~M.} \bibnamefont{Weston}},
	\bibinfo{author}{\bibfnamefont{H.~M.} \bibnamefont{Chrzanowski}},
	\bibinfo{author}{\bibfnamefont{L.~K.} \bibnamefont{Shalm}},
	\bibinfo{author}{\bibfnamefont{V.~B.} \bibnamefont{Verma}},
	\bibinfo{author}{\bibfnamefont{S.~W.} \bibnamefont{Nam}}, \bibnamefont{and}
	\bibinfo{author}{\bibfnamefont{G.~J.} \bibnamefont{Pryde}},
	\bibinfo{journal}{Nature Photonics} \textbf{\bibinfo{volume}{11}},
	\bibinfo{pages}{700} (\bibinfo{year}{2017}).
	
	\bibitem[{\citenamefont{Caves}(1981)}]{caves1981quantum}
	\bibinfo{author}{\bibfnamefont{C.~M.} \bibnamefont{Caves}},
	\bibinfo{journal}{Physical Review D} \textbf{\bibinfo{volume}{23}},
	\bibinfo{pages}{1693} (\bibinfo{year}{1981}).
	
	\bibitem[{\citenamefont{Yurke et~al.}(1986)\citenamefont{Yurke, McCall, and
			Klauder}}]{yurke19862}
	\bibinfo{author}{\bibfnamefont{B.}~\bibnamefont{Yurke}},
	\bibinfo{author}{\bibfnamefont{S.~L.} \bibnamefont{McCall}},
	\bibnamefont{and} \bibinfo{author}{\bibfnamefont{J.~R.}
		\bibnamefont{Klauder}}, \bibinfo{journal}{Physical Review A}
	\textbf{\bibinfo{volume}{33}}, \bibinfo{pages}{4033} (\bibinfo{year}{1986}).
	
	\bibitem[{\citenamefont{Schnabel et~al.}(2010)\citenamefont{Schnabel,
			Mavalvala, McClelland, and Lam}}]{schnabel2010quantum}
	\bibinfo{author}{\bibfnamefont{R.}~\bibnamefont{Schnabel}},
	\bibinfo{author}{\bibfnamefont{N.}~\bibnamefont{Mavalvala}},
	\bibinfo{author}{\bibfnamefont{D.~E.} \bibnamefont{McClelland}},
	\bibnamefont{and} \bibinfo{author}{\bibfnamefont{P.~K.} \bibnamefont{Lam}},
	\bibinfo{journal}{Nature communications} \textbf{\bibinfo{volume}{1}},
	\bibinfo{pages}{121} (\bibinfo{year}{2010}).
	
	\bibitem[{\citenamefont{Abadie et~al.}(2011)\citenamefont{Abadie, Abbott,
			Abbott, Abbott, Abernathy, Adams, Adhikari, Affeldt, Allen, Allen
			et~al.}}]{abadie2011gravitational}
	\bibinfo{author}{\bibfnamefont{J.}~\bibnamefont{Abadie}},
	\bibinfo{author}{\bibfnamefont{B.}~\bibnamefont{Abbott}},
	\bibinfo{author}{\bibfnamefont{R.}~\bibnamefont{Abbott}},
	\bibinfo{author}{\bibfnamefont{T.}~\bibnamefont{Abbott}},
	\bibinfo{author}{\bibfnamefont{M.}~\bibnamefont{Abernathy}},
	\bibinfo{author}{\bibfnamefont{C.}~\bibnamefont{Adams}},
	\bibinfo{author}{\bibfnamefont{R.}~\bibnamefont{Adhikari}},
	\bibinfo{author}{\bibfnamefont{C.}~\bibnamefont{Affeldt}},
	\bibinfo{author}{\bibfnamefont{B.}~\bibnamefont{Allen}},
	\bibinfo{author}{\bibfnamefont{G.}~\bibnamefont{Allen}},
	\bibnamefont{et~al.}, \bibinfo{journal}{Nature Physics}
	\textbf{\bibinfo{volume}{7}}, \bibinfo{pages}{962} (\bibinfo{year}{2011}).
	
	\bibitem[{\citenamefont{Aasi et~al.}(2013)\citenamefont{Aasi, Abadie, Abbott,
			Abbott, Abbott, Abernathy, Adams, Adams, Addesso, Adhikari
			et~al.}}]{aasi2013enhanced}
	\bibinfo{author}{\bibfnamefont{J.}~\bibnamefont{Aasi}},
	\bibinfo{author}{\bibfnamefont{J.}~\bibnamefont{Abadie}},
	\bibinfo{author}{\bibfnamefont{B.}~\bibnamefont{Abbott}},
	\bibinfo{author}{\bibfnamefont{R.}~\bibnamefont{Abbott}},
	\bibinfo{author}{\bibfnamefont{T.}~\bibnamefont{Abbott}},
	\bibinfo{author}{\bibfnamefont{M.}~\bibnamefont{Abernathy}},
	\bibinfo{author}{\bibfnamefont{C.}~\bibnamefont{Adams}},
	\bibinfo{author}{\bibfnamefont{T.}~\bibnamefont{Adams}},
	\bibinfo{author}{\bibfnamefont{P.}~\bibnamefont{Addesso}},
	\bibinfo{author}{\bibfnamefont{R.}~\bibnamefont{Adhikari}},
	\bibnamefont{et~al.}, \bibinfo{journal}{Nature Photonics}
	\textbf{\bibinfo{volume}{7}}, \bibinfo{pages}{613} (\bibinfo{year}{2013}).
	
	\bibitem[{\citenamefont{Mitchell et~al.}(2004)\citenamefont{Mitchell, Lundeen,
			and Steinberg}}]{Mitchell04super}
	\bibinfo{author}{\bibfnamefont{M.~W.} \bibnamefont{Mitchell}},
	\bibinfo{author}{\bibfnamefont{J.~S.} \bibnamefont{Lundeen}},
	\bibnamefont{and} \bibinfo{author}{\bibfnamefont{A.~M.}
		\bibnamefont{Steinberg}}, \bibinfo{journal}{Nature}
	\textbf{\bibinfo{volume}{429}}, \bibinfo{pages}{161} (\bibinfo{year}{2004}).
	
	\bibitem[{\citenamefont{Walther et~al.}(2004)\citenamefont{Walther, Pan,
			Aspelmeyer, Ursin, Gasparoni, and Zeilinger}}]{walther2004broglie}
	\bibinfo{author}{\bibfnamefont{P.}~\bibnamefont{Walther}},
	\bibinfo{author}{\bibfnamefont{J.-W.} \bibnamefont{Pan}},
	\bibinfo{author}{\bibfnamefont{M.}~\bibnamefont{Aspelmeyer}},
	\bibinfo{author}{\bibfnamefont{R.}~\bibnamefont{Ursin}},
	\bibinfo{author}{\bibfnamefont{S.}~\bibnamefont{Gasparoni}},
	\bibnamefont{and}
	\bibinfo{author}{\bibfnamefont{A.}~\bibnamefont{Zeilinger}},
	\bibinfo{journal}{Nature} \textbf{\bibinfo{volume}{429}},
	\bibinfo{pages}{158} (\bibinfo{year}{2004}).
	
	\bibitem[{\citenamefont{Giovannetti et~al.}(2006)\citenamefont{Giovannetti,
			Lloyd, and Maccone}}]{GiovLM06}
	\bibinfo{author}{\bibfnamefont{V.}~\bibnamefont{Giovannetti}},
	\bibinfo{author}{\bibfnamefont{S.}~\bibnamefont{Lloyd}}, \bibnamefont{and}
	\bibinfo{author}{\bibfnamefont{L.}~\bibnamefont{Maccone}},
	\bibinfo{journal}{Phys. Rev. Lett.} \textbf{\bibinfo{volume}{96}},
	\bibinfo{pages}{010401} (\bibinfo{year}{2006}).
	
	\bibitem[{\citenamefont{Bollinger et~al.}(1996)\citenamefont{Bollinger, Itano,
			Wineland, and Heinzen}}]{Boll96optimal}
	\bibinfo{author}{\bibfnamefont{J.~J.} \bibnamefont{Bollinger}},
	\bibinfo{author}{\bibfnamefont{W.~M.} \bibnamefont{Itano}},
	\bibinfo{author}{\bibfnamefont{D.~J.} \bibnamefont{Wineland}},
	\bibnamefont{and} \bibinfo{author}{\bibfnamefont{D.}~\bibnamefont{Heinzen}},
	\bibinfo{journal}{Physical Review A} \textbf{\bibinfo{volume}{54}},
	\bibinfo{pages}{R4649} (\bibinfo{year}{1996}).
	
	\bibitem[{\citenamefont{Lee et~al.}(2002)\citenamefont{Lee, Kok, and
			Dowling}}]{Lee02a}
	\bibinfo{author}{\bibfnamefont{H.}~\bibnamefont{Lee}},
	\bibinfo{author}{\bibfnamefont{P.}~\bibnamefont{Kok}}, \bibnamefont{and}
	\bibinfo{author}{\bibfnamefont{J.~P.} \bibnamefont{Dowling}},
	\bibinfo{journal}{Journal of Modern Optics} \textbf{\bibinfo{volume}{49}},
	\bibinfo{pages}{2325} (\bibinfo{year}{2002}).
	
	\bibitem[{\citenamefont{Okamoto et~al.}(2008)\citenamefont{Okamoto, Hofmann,
			Nagata, O'Brien, Sasaki, and Takeuchi}}]{Okam08beating}
	\bibinfo{author}{\bibfnamefont{R.}~\bibnamefont{Okamoto}},
	\bibinfo{author}{\bibfnamefont{H.~F.} \bibnamefont{Hofmann}},
	\bibinfo{author}{\bibfnamefont{T.}~\bibnamefont{Nagata}},
	\bibinfo{author}{\bibfnamefont{J.~L.} \bibnamefont{O'Brien}},
	\bibinfo{author}{\bibfnamefont{K.}~\bibnamefont{Sasaki}}, \bibnamefont{and}
	\bibinfo{author}{\bibfnamefont{S.}~\bibnamefont{Takeuchi}},
	\bibinfo{journal}{New Journal of Physics} \textbf{\bibinfo{volume}{10}},
	\bibinfo{pages}{073033} (\bibinfo{year}{2008}).
	
	\bibitem[{\citenamefont{Afek et~al.}(2010)\citenamefont{Afek, Ambar, and
			Silberberg}}]{Afek10high}
	\bibinfo{author}{\bibfnamefont{I.}~\bibnamefont{Afek}},
	\bibinfo{author}{\bibfnamefont{O.}~\bibnamefont{Ambar}}, \bibnamefont{and}
	\bibinfo{author}{\bibfnamefont{Y.}~\bibnamefont{Silberberg}},
	\bibinfo{journal}{Science} \textbf{\bibinfo{volume}{328}},
	\bibinfo{pages}{879} (\bibinfo{year}{2010}).
	
	\bibitem[{\citenamefont{Berry et~al.}(2009)\citenamefont{Berry, Higgins,
			Bartlett, Mitchell, Pryde, and Wiseman}}]{Berr09how}
	\bibinfo{author}{\bibfnamefont{D.~W.} \bibnamefont{Berry}},
	\bibinfo{author}{\bibfnamefont{B.~L.} \bibnamefont{Higgins}},
	\bibinfo{author}{\bibfnamefont{S.~D.} \bibnamefont{Bartlett}},
	\bibinfo{author}{\bibfnamefont{M.~W.} \bibnamefont{Mitchell}},
	\bibinfo{author}{\bibfnamefont{G.~J.} \bibnamefont{Pryde}}, \bibnamefont{and}
	\bibinfo{author}{\bibfnamefont{H.~M.} \bibnamefont{Wiseman}},
	\bibinfo{journal}{Physical Review A} \textbf{\bibinfo{volume}{80}},
	\bibinfo{pages}{052114} (\bibinfo{year}{2009}).
	
	\bibitem[{\citenamefont{Pang and Brun}(2014)}]{pang2014}
	\bibinfo{author}{\bibfnamefont{S.}~\bibnamefont{Pang}} \bibnamefont{and}
	\bibinfo{author}{\bibfnamefont{T.~A.} \bibnamefont{Brun}},
	\bibinfo{journal}{Phys. Rev. A.} \textbf{\bibinfo{volume}{90}},
	\bibinfo{pages}{022117} (\bibinfo{year}{2014}).
	
	\bibitem[{\citenamefont{Yuan and Fung}(2015)}]{Haid15optimal}
	\bibinfo{author}{\bibfnamefont{H.}~\bibnamefont{Yuan}} \bibnamefont{and}
	\bibinfo{author}{\bibfnamefont{C.-H.~F.} \bibnamefont{Fung}},
	\bibinfo{journal}{Phys. Rev. Lett.} \textbf{\bibinfo{volume}{115}},
	\bibinfo{pages}{110401} (\bibinfo{year}{2015}).
	
	\bibitem[{\citenamefont{Holevo}(1982)}]{Hole82book}
	\bibinfo{author}{\bibfnamefont{A.~S.} \bibnamefont{Holevo}},
	\emph{\bibinfo{title}{Probabilistic and Statistical Aspects of Quantum
			Theory}} (\bibinfo{publisher}{North-Holland}, \bibinfo{address}{Amsterdam},
	\bibinfo{year}{1982}).
	
	\bibitem[{\citenamefont{Helstrom}(1976)}]{Hels76book}
	\bibinfo{author}{\bibfnamefont{C.~W.} \bibnamefont{Helstrom}},
	\emph{\bibinfo{title}{Quantum Detection and Estimation Theory}}
	(\bibinfo{publisher}{Academic Press}, \bibinfo{address}{New York},
	\bibinfo{year}{1976}).
	
	\bibitem[{\citenamefont{Pang and Jordan}(2017)}]{pang2017optimal}
	\bibinfo{author}{\bibfnamefont{S.}~\bibnamefont{Pang}} \bibnamefont{and}
	\bibinfo{author}{\bibfnamefont{A.~N.} \bibnamefont{Jordan}},
	\bibinfo{journal}{Nature communications} \textbf{\bibinfo{volume}{8}},
	\bibinfo{pages}{14695} (\bibinfo{year}{2017}).
	
	\bibitem[{\citenamefont{Kwiat et~al.}(1999)\citenamefont{Kwiat, Waks, White,
			Appelbaum, and Eberhard}}]{Kwia99ultrabright}
	\bibinfo{author}{\bibfnamefont{P.~G.} \bibnamefont{Kwiat}},
	\bibinfo{author}{\bibfnamefont{E.}~\bibnamefont{Waks}},
	\bibinfo{author}{\bibfnamefont{A.~G.} \bibnamefont{White}},
	\bibinfo{author}{\bibfnamefont{I.}~\bibnamefont{Appelbaum}},
	\bibnamefont{and} \bibinfo{author}{\bibfnamefont{P.~H.}
		\bibnamefont{Eberhard}}, \bibinfo{journal}{Phys. Rev. A}
	\textbf{\bibinfo{volume}{60}}, \bibinfo{pages}{R773} (\bibinfo{year}{1999}).
	
	\bibitem[{\citenamefont{Wang et~al.}(2016)\citenamefont{Wang, Chen, Li, Huang,
			Liu, Chen, Luo, Su, Wu, Li et~al.}}]{Wang16tenphoton}
	\bibinfo{author}{\bibfnamefont{X.-L.} \bibnamefont{Wang}},
	\bibinfo{author}{\bibfnamefont{L.-K.} \bibnamefont{Chen}},
	\bibinfo{author}{\bibfnamefont{W.}~\bibnamefont{Li}},
	\bibinfo{author}{\bibfnamefont{H.-L.} \bibnamefont{Huang}},
	\bibinfo{author}{\bibfnamefont{C.}~\bibnamefont{Liu}},
	\bibinfo{author}{\bibfnamefont{C.}~\bibnamefont{Chen}},
	\bibinfo{author}{\bibfnamefont{Y.-H.} \bibnamefont{Luo}},
	\bibinfo{author}{\bibfnamefont{Z.-E.} \bibnamefont{Su}},
	\bibinfo{author}{\bibfnamefont{D.}~\bibnamefont{Wu}},
	\bibinfo{author}{\bibfnamefont{Z.-D.} \bibnamefont{Li}},
	\bibnamefont{et~al.}, \bibinfo{journal}{Phys. Rev. Lett.}
	\textbf{\bibinfo{volume}{117}}, \bibinfo{pages}{210502}
	(\bibinfo{year}{2016}),
	\urlprefix\url{https://link.aps.org/doi/10.1103/PhysRevLett.117.210502}.
	
	\bibitem[{\citenamefont{Hou et~al.}(2016)\citenamefont{Hou, Zhu, Xiang, Li, and
			Guo}}]{hou2016error}
	\bibinfo{author}{\bibfnamefont{Z.}~\bibnamefont{Hou}},
	\bibinfo{author}{\bibfnamefont{H.}~\bibnamefont{Zhu}},
	\bibinfo{author}{\bibfnamefont{G.-Y.} \bibnamefont{Xiang}},
	\bibinfo{author}{\bibfnamefont{C.-F.} \bibnamefont{Li}}, \bibnamefont{and}
	\bibinfo{author}{\bibfnamefont{G.-C.} \bibnamefont{Guo}},
	\bibinfo{journal}{JOSA B} \textbf{\bibinfo{volume}{33}},
	\bibinfo{pages}{1256} (\bibinfo{year}{2016}).
	
	\bibitem[{\citenamefont{Simon and Mukunda}(1990)}]{simon1990minimal}
	\bibinfo{author}{\bibfnamefont{R.}~\bibnamefont{Simon}} \bibnamefont{and}
	\bibinfo{author}{\bibfnamefont{N.}~\bibnamefont{Mukunda}},
	\bibinfo{journal}{Physics Letters A} \textbf{\bibinfo{volume}{143}},
	\bibinfo{pages}{165} (\bibinfo{year}{1990}).
	
	\bibitem[{\citenamefont{Ahn and Fessler}(2003)}]{ahn2003standard}
	\bibinfo{author}{\bibfnamefont{S.}~\bibnamefont{Ahn}} \bibnamefont{and}
	\bibinfo{author}{\bibfnamefont{J.~A.} \bibnamefont{Fessler}},
	\bibinfo{journal}{EECS Department, The University of Michigan} pp.
	\bibinfo{pages}{1--2} (\bibinfo{year}{2003}).
	
	\bibitem[{\citenamefont{Tsang et~al.}(2011)\citenamefont{Tsang, Wiseman, and
			Caves}}]{Tsang11fundamental}
	\bibinfo{author}{\bibfnamefont{M.}~\bibnamefont{Tsang}},
	\bibinfo{author}{\bibfnamefont{H.~M.} \bibnamefont{Wiseman}},
	\bibnamefont{and} \bibinfo{author}{\bibfnamefont{C.~M.} \bibnamefont{Caves}},
	\bibinfo{journal}{Physical review letters} \textbf{\bibinfo{volume}{106}},
	\bibinfo{pages}{090401} (\bibinfo{year}{2011}).
	
\end{thebibliography}
\end{document}